\PassOptionsToPackage{final}{hyperref}

\documentclass[draft]{agujournal2019}
\usepackage{url} 
\usepackage{lineno}
\usepackage[inline]{trackchanges} 
\usepackage{soul}
\usepackage{natbib}
\usepackage{amsmath}
\usepackage[colorlinks=true, linkcolor=blue, urlcolor=blue, citecolor=blue]{hyperref}
\usepackage{trackchanges}

\draftfalse

\journalname{JGR: Planets}

\begin{document}

\newcommand {\pt}{\frac{\partial}{\partial t}}
\newcommand {\pat}[1]{\frac{\partial #1}{\partial t}}
\newcommand {\upx}{\frac{u}{r} \frac{\partial}{\partial \phi}}
\newcommand {\vpr}{v \frac{\partial}{\partial r}}
\newcommand {\wpz}{w \frac{\partial}{\partial z}}
\newcommand {\px}[1]{\frac{1}{r}\frac{\partial #1}{\partial \phi}}
\newcommand {\pr}[1]{\frac{\partial #1}{\partial r}}
\newcommand {\pz}[1]{\frac{\partial #1}{\partial z}}
\newcommand {\PV}{\mathrm{PV}}
\newcommand {\tu}{\tilde{u}}
\newcommand {\tv}{\tilde{v}}
\newcommand {\tw}{\tilde{w}}
\newcommand {\tb}{\tilde{b}}
\newcommand {\fre}{\xi}

\title{Symmetric instability in a Boussinesq fluid on a rotating planet}

\authors{Yaoxuan Zeng\affil{1} and Malte F. Jansen\affil{1}}

\affiliation{1}{Department of the Geophysical Sciences, The University of Chicago, Chicago, IL 60637, USA}

\correspondingauthor{Yaoxuan Zeng}{yxzeng@uchicago.edu}

\begin{keypoints}
\item Symmetric instability with arbitrary background shear and stratification is studied with linear instability analysis
\item Three types of instability criteria are explored with theoretical analysis and numerical simulations
\item In icy moon oceans with low Rossby number flows slantwise convection occurs if and only if stratification is unstable along rotation axis
\end{keypoints}

\begin{abstract} 
Symmetric instability has broad applications in geophysical and planetary fluid dynamics. It plays a crucial role in the formation of mesoscale rainbands at mid-latitudes on Earth, instability in the ocean's mixed layer, and slantwise convection on gas giants and icy moon oceans. Here, we apply linear instability analysis to an arbitrary zonally symmetric Boussinesq flow on a rotating spherical planet, with applicability to icy moon oceans. We divide the instabilities into three types: (1) gravitational instability, occurring when stratification is unstable along angular momentum surfaces, (2) inertial instability, occurring when angular momentum shear is unstable along buoyancy surfaces, and (3) a mixed symmetric instability, occurring when neither of the previous conditions are fulfilled, but the potential vorticity has the opposite sign to planetary rotation. We note that $N^2<0$ where $N$ is the Brunt--Väisälä frequency---a typical criterion used to trigger convective adjustment in global ocean models---is neither necessary nor sufficient for instability. Instead, $b_z \sin{\theta_0}<0$, where $b_z$ is the stratification along the planetary rotation axis and $\theta_0$ is the local latitude, is always sufficient for instability and also necessary in the low Rossby number limit. In this limit, relevant for deep convection in icy moon oceans, the most unstable mode is slantwise convection parallel to the planetary rotation axis. This slantwise convection differs from the parameterized convection in existing general circulation models, whose convection schemes parameterize convection in the direction of gravity. Our results suggest that convection schemes in global ocean models must be revised before being applied to icy moon oceans.
\end{abstract}

\section*{Plain Language Summary} 
Flows on rotating planets can become unstable because of the combined effects of rotation and density stratification, a phenomenon known as symmetric instability. This instability shapes the flow patterns seen in planetary atmospheres and oceans. In our study, we use theoretical analysis and numerical simulations to study the instability criteria and the most unstable modes for axisymmetric flows on rotating planets. For flows strongly affected by rotation—typical of icy moon oceans—instability occurs if and only if the stratification is unstable along the rotation axis, leading to slantwise convection that aligns with the planetary rotation axis. This result is consistent with the rotation-aligned structures in global numerical simulations for the icy moon oceans. Additionally, it suggests that the traditional convection parameterization in ocean models, which only considers unstable stratification and heat transport in the direction of gravity, is not applicable for the icy moon oceans.

\section{Introduction}\label{sec:intro}

Symmetric instability describes the instability of axisymmetric flow. In an axisymmetric rotating fluid with background density stratification and angular momentum gradient, if a fluid parcel is perturbed from its origin, two restoring forces come into play: the buoyancy force and the inertial acceleration (Coriolis and centrifugal forces). These forces individually may result in gravitational instability or inertial instability. Moreover, even when the fluid is both gravitationally and inertially stable, the combined effects of the two force anomalies may result in symmetric instability \citep{solberg1936mouvement,hoiland1941,hoskins_role_1974,haine_gravitational_1998}.

Symmetric instability has broad applications in geophysical fluid dynamics. On Earth, it is relevant to the formation of mesoscale rainbands in the midlatitude atmosphere \citep[e.g.,][]{emanuel1983lagrangian,emanuel1985convective} and instability in the ocean's mixed layer \citep[e.g., ][]{straneo_idealized_2002,callies_baroclinic_2018}. Beyond Earth, symmetric instability is closely tied to slantwise convection in the atmospheres of gas giants \citep[e.g.,][]{stone1967application,busse_thermal_1970,stone_symmetric_1971,walton_viscous_1975,oneill_slantwise_2016} and in the oceans of their icy satellites \citep[e.g.,][]{soderlund2019ocean,ashkenazy2021dynamic,kang2022does,bire2022exploring,zeng2024effect}, where convection is tilted along angular momentum surfaces.

Symmetric instability is often studied under certain assumptions about the background field. In most literature, it is examined in inertially stable ($\eta/f>0$) fluids with stable gravitational stratification ($N^2>0$), where $N^2=- (g/\rho) (\partial \rho/\partial R)$, $\eta=f-(1/R)(\partial u/\partial \theta)$ is the absolute vorticity, $g$ is gravity, $\rho$ is density, $u$ is the zonal component of the velocity, $f=2\Omega \sin{\theta}$ is the component of planetary rotation parallel to gravity, $\theta \in (-\pi/2,\pi/2]$ is the latitude, $R$ is the planetary radius, and $\partial/\partial R$ denotes the derivative along the gravitational direction. Under these assumptions, instability occurs when the planetary vorticity has the opposite sign to the potential vorticity. In the Northern Hemisphere, symmetric instability arises when the potential vorticity, defined as $q=\nabla b \cdot ( 2\mathbf{\Omega} + \nabla \times \mathbf{v})$, is negative, where $\mathbf{\Omega}$ represents planetary rotation, $\mathbf{v}$ is the three-dimensional (3-D) relative velocity in the rotating frame, and $b$ is the buoyancy \citep{eliassen1951slow,ooyama1966stability,hoskins_role_1974,stevens1983symmetric}. Although assuming $N^2>0$ is reasonable for most regions of Earth's atmosphere and ocean, this condition may not hold on other planetary bodies such as icy moon oceans and gas giant atmospheres. Studies on rotating convective instability have examined scenarios with $N^2<0$, and have found that rotating systems can remain stable even when $N^2<0$ \citep{flasar_turbulent_1978,hathaway_convective_1979}. In \cite{flasar_turbulent_1978} and \cite{hathaway_convective_1979}, the background zonal shear is assumed to be purely vertical. However, meridional shears resulting from eddy angular momentum transport \citep[e.g.,][]{busse_thermal_1970,aurnou2007effects,soderlund2019ocean,zeng2021ocean} are likely significant in the oceans of icy moons and the atmospheres of gas giants, thereby influencing the instability criteria.

Beyond the background field, various approximations have been employed in studies of symmetric instability. In early studies, the traditional approximation \citep{gerkema2008geophysical} has been commonly applied, wherein the problem is examined on an $f$-plane that considers only the component of planetary rotation parallel to gravity. In this framework, the instability criteria for gravitationally and inertially stable fluids can be expressed as $\mathrm{Ri}<\eta/f$, the same as negative PV in the Northern Hemisphere \citep[c.f.][]{hoskins_role_1974,haine_gravitational_1998}, where $\mathrm{Ri}=N^2/(\partial u/\partial R)^2$ is the Richardson number. However, the traditional approximation is not valid for planets with deep fluid layers, such as the atmospheres of gas giants and the oceans of icy moons, for flows with strong vertical motion, or for near-equatorial flows where $f\to 0$ \citep{gerkema2008geophysical}. Later studies have considered a ``tilted $f$-plane'' that takes into account both vertical and horizontal components of the planetary rotation, $\mathbf{f}= f \mathbf{e_R} + \tilde{f} \mathbf{e_\theta}$ where $\tilde{f}=2\Omega \cos{\theta}$, $\mathbf{e_R}$ denotes the upwards direction parallel to gravity, and $\mathbf{e_\theta}$ denotes the latitudinal direction defined to point northward \citep[see Figure~\ref{fig1:coordinate};][]{sun_unsymmetrical_1995,straneo_idealized_2002,fruman_symmetric_2008,itano_symmetric_2009,jeffery_effect_2009}. \cite{sun_unsymmetrical_1995} concluded that as latitude decreases, i.e., the angle between planetary rotation and gravity becomes larger, the effect of $\tilde{f}$ in modulating the maximum growth rate of symmetric instability increases. \cite{itano_symmetric_2009} found that the parameter regime for the occurrence of symmetric instability is less sensitive to $\tilde{f}$ when $\mathrm{Ri}>0.25$ and $\eta/f > 1$, but is considerably influenced in other regions.

The instabilities discussed above are generally difficult to resolve in global ocean simulations due to their small spatial scales. In studies of icy moon oceans, two main approaches have been used. The first involves direct numerical simulations in parameter regimes that differ from those of actual icy moon oceans, with asymptotic scaling laws developed to extrapolate the results \citep{soderlund2014ocean,gastine2016scaling,soderlund2019ocean,amit2020cooling,kvorka2022numerical,cabanes2024zonostrophic,bouffard2025seafloor}. Parameters are chosen such that convective instabilities can be explicitly resolved, but results often need to be extrapolated by several orders of magnitude to infer the circulation on icy moons, raising concerns about their reliability in representing real ocean dynamics. The second approach uses sub-grid parameterizations to represent unresolved processes (including convection), enabling simulations under realistic icy moon conditions \citep{kang2022role,kang2022does,kang2023modulation,zeng2024effect}. However, most parameterizations are designed for Earth-like flows, and whether they are applicable to other planets remains unclear. Here, we discuss symmetric instability conditions and unstable mode properties over a wide planetary parameter regime with slowly varying, but otherwise arbitrary zonal velocity shear and density stratification. We apply a local linear instability analysis to the adiabatic, inviscid, non-hydrostatic Boussinesq equations formulated in a cylindrical coordinate system, which is the most natural choice for analyzing fast-rotating planets, such as icy moon oceans, where motions are largely aligned with the rotation axis \citep{ashkenazy2021dynamic,bire2022exploring}. We retain full Coriolis force and metric terms that arise from the cylindrical geometry of the coordinate system. We linearize around a zonally symmetric background state in hydrostatic and gradient wind balance, without imposing additional assumptions on this state. In particular, we do not require the background stratification to be stable in the gravitational direction ($N^2>0$), which, as we show, is in general neither a necessary nor sufficient condition for stability. We consider zonally-symmetric perturbations, which allow for gravitational, inertial, and mixed symmetric instabilities, as discussed in this paper, while excluding baroclinic instabilities associated with nonzero zonal wavenumbers \citep{stone1966non}. In Section~\ref{sec:linear_analysis}, the theoretical analysis for linear instability criteria, the most unstable mode, and the maximum growth rate are discussed. With hindsight, the instability criteria derived here are essentially the Boussinesq analog of the Solberg-Høiland criteria \citep{solberg1936mouvement,hoiland1941} as formulated in \cite{ogilvie2019internal}. Section~\ref{sec:numerical_simulations} presents the numerical simulation results for comparison with the theoretical analysis. Section~\ref{sec:discussions} provides discussions. Section~\ref{sec:conclusions} provides concluding remarks.

\section{Linear instability analysis}\label{sec:linear_analysis}

\subsection{Instability criteria}\label{subsec:stability}

\begin{figure}[t]
\centering
\includegraphics[width=0.35\linewidth]{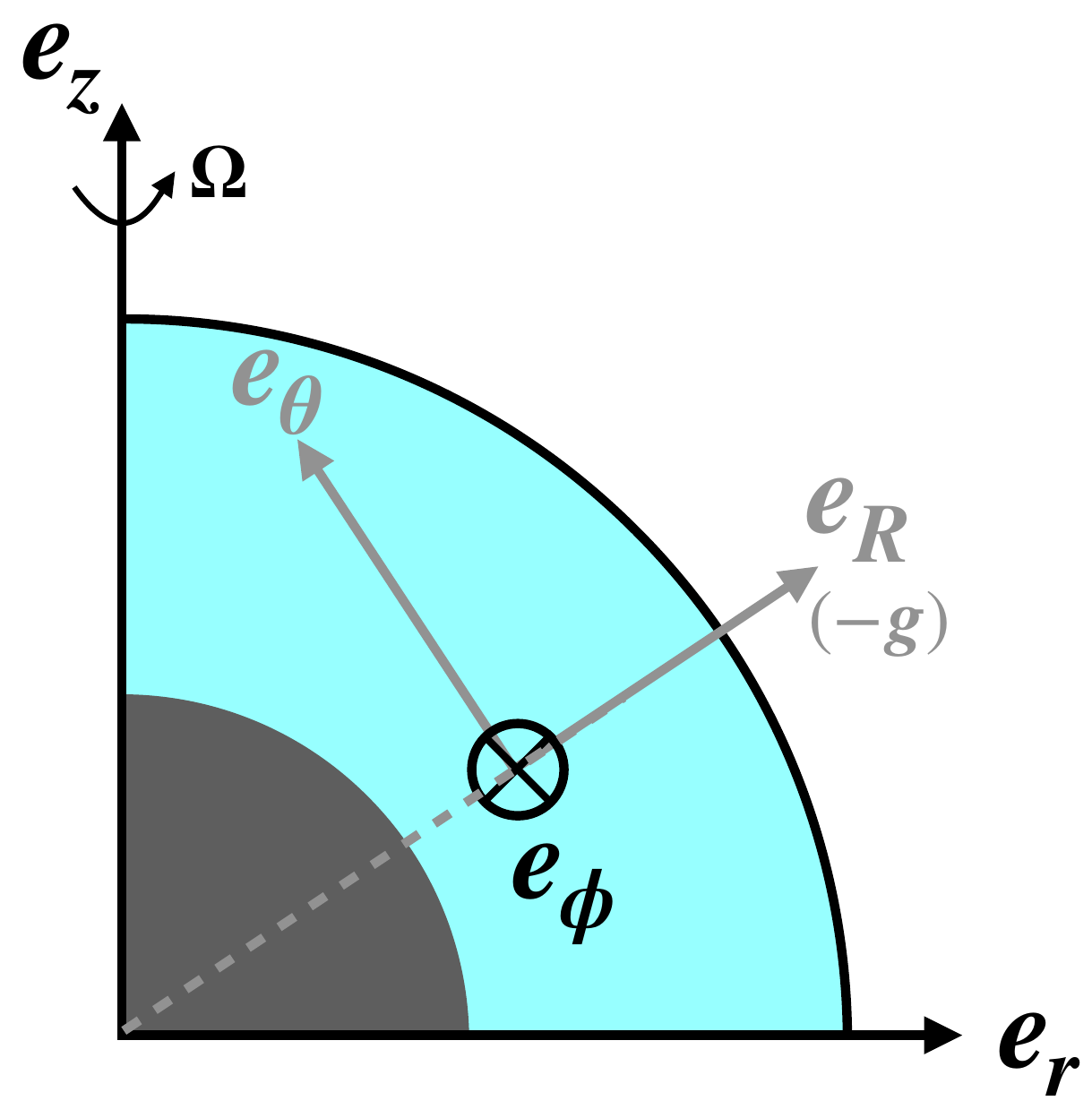}
\caption{Sketch of the coordinate system. Black coordinates show the cylindrical coordinate system applied in this paper ($\mathbf{e_r}$, $\mathbf{e_\phi}$, $\mathbf{e_z}$), where $\mathbf{e_z}$ is the planetary rotation axis and the origin locates at the center of the planet. Grey coordinates show the spherical coordinate system ($\mathbf{e_\phi}$, $\mathbf{e_\theta}$, $\mathbf{e_R}$), where $\mathbf{e_R}$ is opposite to gravity. Note that the zonal direction $\mathbf{e_\phi}$ is the same in both cylindrical and spherical coordinates. \label{fig1:coordinate}}
\end{figure}

The adiabatic, inviscid, nonhydrostatic equations for a Boussinesq fluid in a cylindrical coordinate are the three components of the Navier-Stokes equations:

\begin{equation}\label{eq4:u}
    \left( \pt + \upx + \vpr + \wpz \right)u + 2\Omega v + \frac{uv}{r} = -\px{\Phi},
\end{equation}

\begin{equation}\label{eq5:v}
    \left( \pt + \upx + \vpr + \wpz \right)v - 2\Omega u - \frac{u^2}{r} = -\pr{\Phi} + b\cos{\theta},
\end{equation}

\begin{equation}\label{eq6:w}
    \left( \pt + \upx + \vpr + \wpz \right)w = -\pz{\Phi} + b\sin{\theta},
\end{equation}

\noindent the buoyancy equation:

\begin{equation}\label{eq7:b}
    \left( \pt + \upx + \vpr + \wpz \right)b = 0,
\end{equation}

\noindent and the mass continuity equation:

\begin{equation}\label{eq8:cont}
    \px{u} + \pr{v} + \pz{w} + \frac{v}{r} = 0,
\end{equation}

\noindent where $r$ is the radial cylindrical direction, increasing outwards perpendicular to the rotation axis of the planet, $\phi$ is the longitude, denoting the zonal (azimuthal) direction, and $z$ is the vertical (rotational) direction (Figure~\ref{fig1:coordinate}). $u$, $v$, $w$ are velocity components in the $\mathbf{e_\phi}$, $\mathbf{e_r}$, and $\mathbf{e_z}$ directions, respectively. $\Phi=p'/\rho_0$, where $p'$ is the pressure anomaly and $\rho_0$ is the reference density. $b = -g\rho'/\rho_0$ is buoyancy, where $\rho'$ is the density anomaly \citep[c.f. ][]{vallis2017atmospheric}.

We consider an axisymmetric state around the planetary rotation axis where all variables are invariant in the zonal direction ($\partial/\partial \phi = 0$, i.e., zonally symmetric), and assume a background state with an arbitrary background zonal flow $\overline{u}(r,z)$ and buoyancy field $\overline{b}(r,z)$ that are invariant in time: $b = \overline{b}(r,z) + b'(r,z,t)$, $\Phi = \overline{\Phi}(r,z) + \Phi'(r,z,t)$, $u = \overline{u}(r,z) + u'(r,z,t)$, $v=v'(r,z,t)$, $w=w'(r,z,t)$. Assuming the perturbations are small, the zeroth-order balance reveals the gradient wind balance and hydrostatic balance for the background state:

\begin{equation}\label{eq9:thermalwind1}
    -2 \Omega \overline{u} - \frac{\overline{u}^2}{r} = -\pr{\overline{\Phi}} + \overline{b} \cos{\theta},
\end{equation}

\begin{equation}\label{eq10:thermalwind2}
    0 = -\pz{\overline{\Phi}} + \overline{b} \sin{\theta}.
\end{equation}

With

\begin{equation}
    \overline{f} = 2\Omega + 2 \overline{\omega}, 
\end{equation}

\noindent being a modified Coriolis parameter where $\overline{\omega} \equiv \overline{u}/r$ is the angular velocity of the background zonal flow, Equations~\ref{eq9:thermalwind1}~\&~\ref{eq10:thermalwind2} yield

\begin{equation}\label{eq:gradientwind}
    \overline{f} \pz{\overline{u}} = \frac{\partial \overline{b}}{\partial r} \sin{\theta} - \frac{\partial \overline{b}}{\partial z} \cos{\theta}.
\end{equation}

\noindent Equation~\ref{eq:gradientwind} is similar to the thermal wind balance in a rapidly rotating (geostrophic) fluid \citep{kaspi2009deep,bire2022exploring}, but here $\overline{f}$ is modified by the background flow, with the effects of centrifugal force incorporated, thus giving the shear of a flow in gradient-wind balance rather than geostrophic balance.

In the first-order balance, we obtain linearized perturbation equations (Equations~\ref{eqa6:ulinear}-\ref{eqa10:contlinear}). In this paper, we focus on local instability analysis, with the length scales of the perturbations small compared to the planetary radius. We therefore look for plane-wave solutions, where all variables are proportional to $\exp{(ik_r r + ik_z z - i \fre t)}$, where $\fre$ is the angular frequency, and $k_r$ and $k_z$ are wavenumbers in the radial and vertical directions, respectively. We assume that variations in the background field are small across the perturbation length scale (Equation~\ref{eq:background_approx_condition}), allowing us to evaluate them locally at $(r=r_0, z=z_0)$, with $k_r$ and $k_z$ treated as constants. Under these assumptions, we have the dispersion relation (see \ref{app:dispersion} for detailed derivation)

\begin{equation}\label{eq18:dispersion}
    \fre^2 = \frac{b_z \sin{\theta_0} k_r^2 + (f_0 M_r + b_r \cos{\theta_0}) k_z^2 - 2 b_r \sin{\theta_0} k_r k_z}{k_r^2 + k_z^2},
\end{equation}

\noindent where $b_z \equiv \partial \overline{b}/\partial z|_{(r_0,z_0)}$ and $b_r \equiv \partial \overline{b}/\partial r|_{(r_0,z_0)}$ are stratification, $f_0 \equiv 2\Omega + 2 \overline{\omega}$ is the modified Coriolis parameter, and $M_r \equiv (1/r_0)(\partial \overline{m}/\partial r)|_{(r_0,z_0)} = f_0 + r_0(\partial \overline{\omega}/\partial r)|_{(r_0,z_0)}$ describes the radial gradient of the background angular momentum $\overline{m} = (\Omega+\overline{\omega}) r^2$, all evaluated at $(r_0, z_0)$. $\theta_0=\arctan{(z_0/r_0)}$ is the local latitude.

Stability requires that the frequency $\fre$ has no imaginary part, which means the right-hand-side of Equation~\ref{eq18:dispersion} is positive definite. Specifically, this requires the quadratic function $b_z \sin{\theta_0} k_r^2 + (f_0 M_r + b_r \cos{\theta_0}) k_z^2 - 2 b_r \sin{\theta_0} k_r k_z>0$ for all $k_r$ and $k_z$. Consequently, the stability matrix

\begin{equation*}
    \left( \begin{matrix}
        b_z \sin{\theta_0} & -b_r \sin{\theta_0} \\
        -b_r \sin{\theta_0} & f_0 M_r + b_r \cos{\theta_0} \\
    \end{matrix}\right)
\end{equation*}

\noindent must be positive definite. This stability matrix is essentially the same as the one derived in previous studies on symmetric instability on a tilted $f$-plane \citep[e.g.,][]{itano_symmetric_2009} although here it is expressed in cylindrical coordinates instead of spherical coordinates to emphasize the symmetry in flows that are strongly rotationally constrained, and the Coriolis parameter $f_0$ is modified by the angular velocity of the background flow $\overline{\omega}$. It also closely resembles the corresponding expression for a compressible atmosphere in cylindrical coordinates \citep{ogilvie2019internal}. The necessary and sufficient conditions for instability are that either the trace or determinant of this matrix be negative, i.e., either

\begin{equation}\label{eq:trace}
    b_z \sin{\theta_0} + f_0 M_r + b_r \cos{\theta_0} < 0,
\end{equation}

\noindent or

\begin{equation}\label{eq:det}
    b_z \sin{\theta_0}(f_0 M_r + b_r \cos{\theta_0}) - b_r^2 \sin^2{\theta_0} < 0.
\end{equation}

Equation~\ref{eq:det} can also be expressed in terms of a condition for the background potential vorticity. The background potential vorticity is

\begin{equation}\label{eq:pvdef}
    q_0 = \nabla \overline{b} \cdot (2 \mathbf{\Omega} + \nabla \times \mathbf{\overline{v}}) = M_rb_z-M_zb_r = M_r b_z - \frac{b_r}{f_0}(b_r \sin{\theta_0} - b_z \cos{\theta_0}),
\end{equation}

\noindent where $M_z\equiv (1/r_0)(\partial \overline{m}/\partial z)|_{(r_0,z_0)}= r_0(\partial \overline{\omega}/\partial z)|_{(r_0,z_0)}$ describes the vertical gradient of the background angular momentum, and the background gradient wind shear (Equation~\ref{eq:gradientwind}) is applied. We can hence rewrite Equation~\ref{eq:det} as

\begin{equation}\label{eq:cret3}
    q_0 f_0 \sin{\theta_0} < 0,
\end{equation}

\noindent i.e., instability occurs if the potential vorticity has the opposite sign to the generalized background planetary vorticity. This instability criterion is commonly referred to as the symmetric instability criterion \citep[e.g.,][]{hoskins_role_1974,haine_gravitational_1998}, although it should be noted that in general, Equation~\ref{eq:cret3} by itself is a sufficient but not necessary condition for instability.

\subsection{Most unstable mode and growth rate}\label{subsec:unstable_mode}

When $\fre$ is imaginary, the perturbation fields will grow exponentially. Note that $\fre^2$ is a real number (Equation~\ref{eq18:dispersion}); therefore, $\fre$ is either purely real (corresponding to neutral wave solutions) or purely imaginary (corresponding to exponentially growing or decaying solutions). By substituting $\fre = i \sigma$ in the dispersion relation (Equation~\ref{eq18:dispersion}), the e-folding growth rate $\sigma$ of the unstable mode can be expressed as

\begin{equation}\label{eq:growthrate}
    \sigma = \left( -\frac{b_z \sin{\theta_0} \tan^2{\delta} +2b_r \sin{\theta_0} \tan{\delta} + f_0 M_r + b_r \cos{\theta_0}}{1+ \tan^2{\delta}} \right)^{1/2},
\end{equation}

\noindent where $\delta \in (-\pi/2,\pi/2]$ is the angle between the unstable mode and $\mathbf{e_r}$ so that $\tan{\delta} = -k_r/k_z$. Equation~\ref{eq:growthrate} indicates that for a given background field and latitude, the growth rate is only a function of $\tan \delta$, i.e., the direction of the unstable mode, but not the magnitude of the wavenumber, i.e., the size of the mode. The result that the growth rate depends only on the direction of the mode is consistent with previous studies of symmetric instability on an $f$-plane \citep[e.g.,][]{ooyama1966stability} and a tilted $f$-plane \citep[e.g.,][]{sun_unsymmetrical_1995}. By calculating the derivatives of the function $\sigma(\tan{\delta})$ and performing some algebra, we find that the maximum growth rate is obtained when 

\begin{equation}\label{eq:most_unstable_mode}
    \tan{\delta_m} = \frac{b_z \sin{\theta_0} - f_0 M_r - b_r \cos{\theta_0} - \left[ ( f_0 M_r + b_r \cos{\theta_0} - b_z \sin{\theta_0})^2 + 4 b_r^2 \sin^2{\theta_0} \right]^{1/2}}{2b_r \sin{\theta_0}}.
\end{equation}

It should be noted that $\delta_m$ also characterizes the net heat and zonal momentum transports in the $r$-$z$ plane by the most unstable mode, and their directions align with the mode itself (see \ref{app:energy}). Substituting Equation~\ref{eq:most_unstable_mode} into Equation~\ref{eq:growthrate}, we obtain the maximum growth rate as

\begin{equation}\label{eq:max_growth_rate}
    \sigma_{m} = \left( -\frac{ f_0 M_r + b_r \cos{\theta_0} + b_z \sin{\theta_0} - \sqrt{(f_0 M_r + b_r \cos{\theta_0}-b_z \sin{\theta_0})^2+4 b_r^2 \sin^2{\theta_0}}}{2}\right)^{1/2}.
\end{equation}

\subsection{Instability diagram}\label{subsec:stability_diagram}

In this section, we discuss the instability criteria assuming $b_r \sin{\theta_0} \ne 0$. The special case where $b_r \sin{\theta_0}=0$ is discussed in \ref{app:brsin}, where it is shown that the most unstable mode in this case always aligns with the radial ($r$) or vertical ($z$) directions. The instability criteria for fluid parcels perturbed purely along the $\mathbf{e_r}$ and $\mathbf{e_z}$ directions are $f_0 M_r + b_r \cos{\theta_0}<0$ and $b_z \sin{\theta_0}<0$, respectively, which are derived by solving $\fre^2<0$ (Equation~\ref{eq18:dispersion}) when $\delta=0$ and $\delta=\pi/2$, respectively. It is convenient to define a 2-D nondimensionalized phase space based on the stability in these two orthogonal directions:

\begin{equation}\label{eq:definexy}
    x = \frac{b_z \sin{\theta_0}}{|b_r \sin{\theta_0}|}, \ \ y = \frac{f_0 M_r + b_r \cos{\theta_0}}{|b_r \sin{\theta_0}|}.
\end{equation}

\begin{figure}[b!]
\centering
\includegraphics[width=1.0\linewidth]{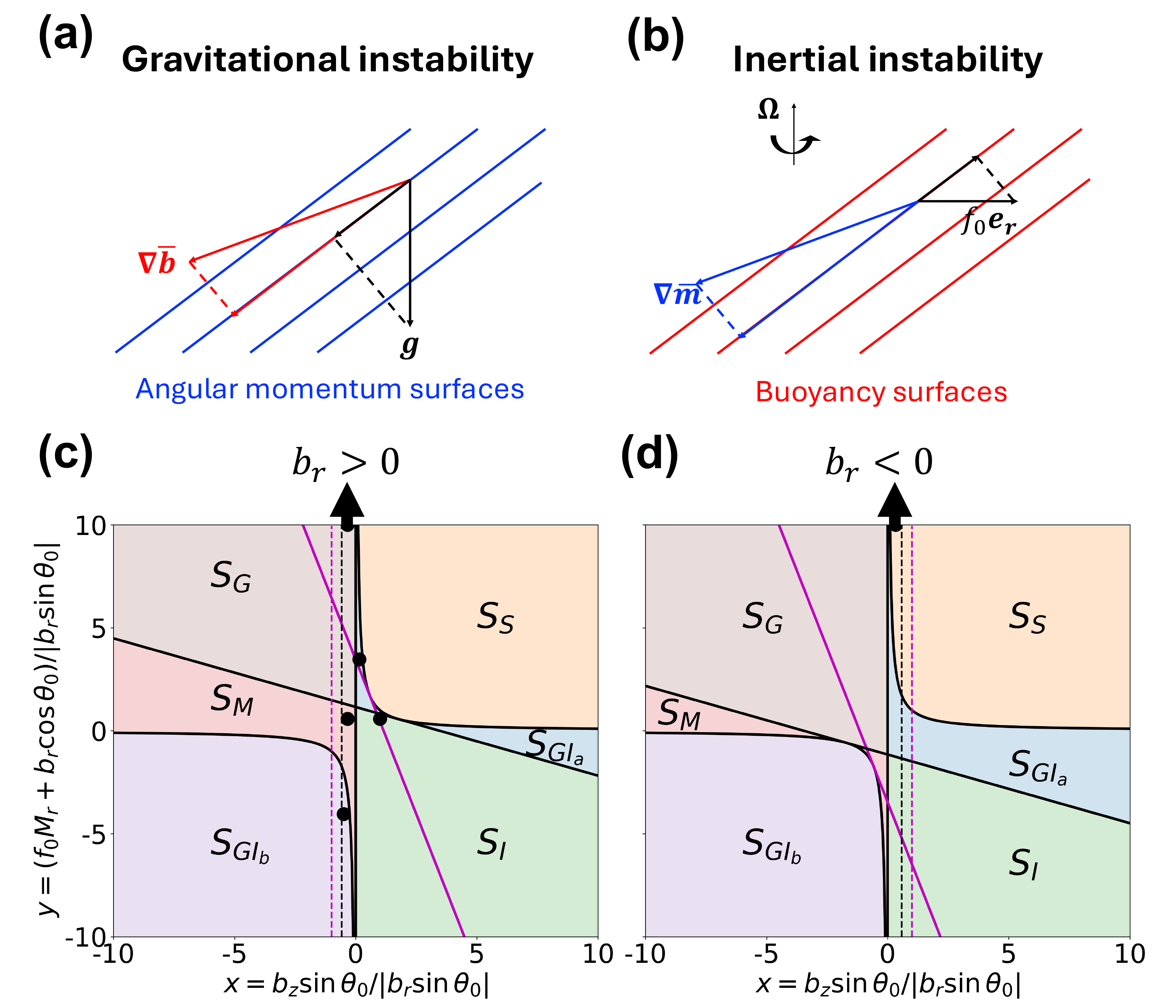}
\caption{Instability diagrams for $\theta_0$=60$^\circ$. (a) and (b) sketch gravitational and inertial instabilities, respectively (see text for details). (c) and (d) show the instability regime diagram with $b_r>0$ in (c) and $b_r<0$ in (d). Subscripts in the labels indicate different types of instability: $G$ for gravitational instability (Equation~\ref{eq:gravitationalvec}), $I$ for inertial instability (Equation~\ref{eq:inertialvec}), $M$ for mixed symmetric instability (Equation~\ref{eq:cret3}), and $S$ for the stable regime. The $GI$ regime is both gravitationally and inertially unstable, with $GI_a$ corresponding to $q_0 f_0 \sin{\theta_0}<0$ and $GI_b$ to $q_0 f_0 \sin{\theta_0}>0$. Dots in (c) and (d) mark simulation parameters listed in Table~\ref{tab:simulations}. In simulations $s_G$ and $s_S$, the $y$-axis value is 1.15 × 10$^4$, which lies outside the plotted range. Black solid lines in (c) and (d) indicate the criteria for gravitational, inertial, and mixed symmetric instabilities (Equations~\ref{eq:nondim_grav}-\ref{eq:nondim-PVcriteria}). The dashed lines indicate $N^2=0$, with regions to the left corresponding to $N^2<0$. The magenta lines indicate the criterion for $\theta_0 = 30^\circ$ for comparison. \label{fig2:instability}}
\end{figure}

The stability matrix then becomes

\begin{equation*}
    \left( \begin{matrix}
        x & -1 \\
        -1 & y \\
    \end{matrix}\right),
\end{equation*}

\noindent and positive definiteness requires that 

\begin{equation}\label{eq:nondim-xcriteria}
    x+y>0, \ \mathrm{and} \ xy>1,
\end{equation}

\noindent which means only the region above the hyperbolic $xy=1$ in the first quadrant ($x>0$, $y>0$) is stable (Figure~\ref{fig2:instability}c~\&~d).

\subsection{Gravitational instability, inertial instability, and mixed symmetric instability}\label{subsec:physical_stability}

In a rotating fluid with background stratification and shear, two restoring forces, associated with gravity and rotation, act when a fluid parcel is displaced: buoyancy force and inertial acceleration. If a fluid parcel is displaced along a constant angular momentum surface, the inertial acceleration anomaly is zero and the only restoring force is the buoyancy force anomaly. As a result, pure gravitational instability can occur when the stratification is unstable along constant angular momentum surfaces. Similarly, if the displacement of the fluid parcel is along a constant buoyancy surface, the only restoring force is the inertial acceleration anomaly, hence pure inertial instability can occur when the angular momentum shear is unstable along constant buoyancy surfaces. Here, unstable stratification indicates buoyancy decreases in the opposite direction to gravity, and unstable angular momentum shear indicates the angular momentum gradient is opposite to the background planetary angular momentum gradient, $f_0 \mathbf{e_r}$ (Figure~\ref{fig2:instability}a~\&~b). Therefore, gravitational instability can occur when

\begin{equation}\label{eq:gravitationalvec}
    (\nabla \overline{b} \cdot \mathbf{e_M}) (-\mathbf{g} \cdot \mathbf{e_M}) = g(M_r b_z - M_z b_r)(M_r \sin{\theta_0} - M_z \cos{\theta_0}) < 0,
\end{equation}

\noindent where $\mathbf{e_M} = -M_z \mathbf{e_r} + M_r \mathbf{e_z}$ is the direction along a constant angular momentum surface. Inertial instability can occur when 

\begin{equation}\label{eq:inertialvec}
    (\nabla \overline{m} \cdot \mathbf{e_b}) (f_0 \mathbf{e_r} \cdot \mathbf{e_b}) = f_0 b_z r_0 (M_r b_z - M_z b_r) < 0,
\end{equation}

\noindent where $\mathbf{e_b} = -b_z \mathbf{e_r} + b_r \mathbf{e_z}$ is the direction along a constant buoyancy surface. Using the gradient wind shear (Equation~\ref{eq:gradientwind}), and the definition of $x$ and $y$ (Equation~\ref{eq:definexy}), the gravitational instability criterion (Equation~\ref{eq:gravitationalvec}) becomes

\begin{equation}\label{eq:nondim_grav}
   [y + \cot^2{\theta_0} x - 2 |\cot{\theta_0}| \mathrm{sgn}(b_r) ](xy-1) < 0,
\end{equation}

\noindent where $\mathrm{sgn}$ denotes the sign function, and the inertial instability criterion (Equation~\ref{eq:inertialvec}) becomes

\begin{equation}\label{eq:nondim_iner}
    x(xy-1) < 0.
\end{equation}

Comparison with the results derived in Section~\ref{subsec:stability} shows that neither the gravitational nor inertial instability criteria are necessary for instability. Instability can still occur as a result of gravitational and inertial force anomalies when $q_0 f_0 \sin{\theta_0}<0$, or equivalently, 

\begin{equation}\label{eq:nondim-PVcriteria}
    xy - 1 < 0.
\end{equation}

\noindent We therefore refer to this instability (Equation~\ref{eq:nondim-PVcriteria}) when the flow is both gravitationally stable and inertially stable as a mixed symmetric instability.

\begin{figure}[t!]
\centering
\includegraphics[width=1.0\linewidth]{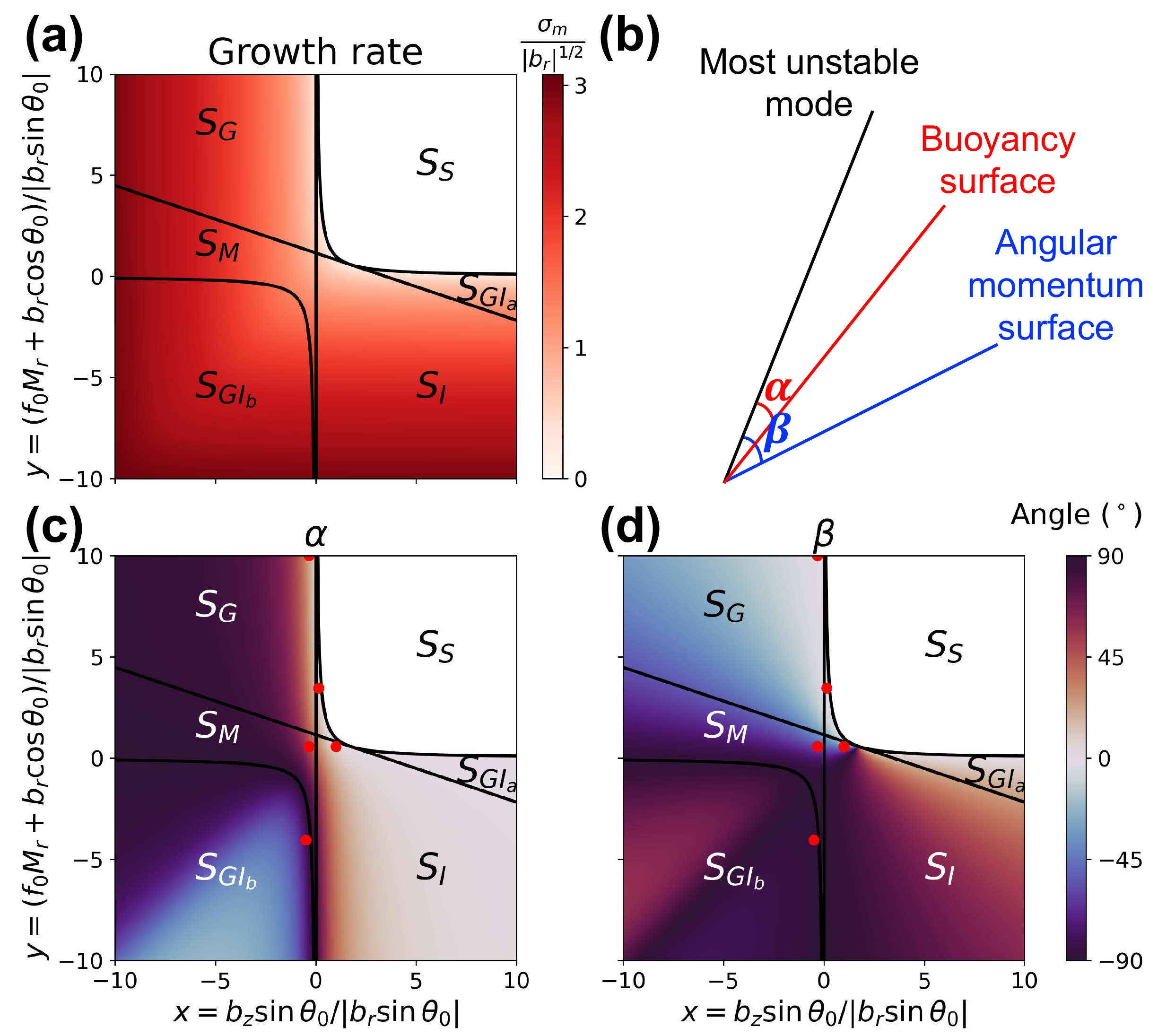}
\caption{Growth rate and orientation of the most unstable mode with positive $b_r$ here shown for latitude $\theta_0$=60$^\circ$. (a): growth rate of the most unstable mode, nondimensionalized with $|b_r|^{1/2}$: $\sigma_m/|b_r|^{1/2} = ((-x-y+((x-y)^2+4)^{1/2})|\sin{\theta_0}|/2)^{1/2}$. (b) defines the angles of the most unstable mode relative to the buoyancy ($\alpha$) and the angular momentum surface ($\beta$), shown in (c) and (d), respectively. In (a), (c), and (d), the dots, solid lines, and labels are the same as in Figure~\ref{fig2:instability}c. Note that $+90^\circ$ and $-90^\circ$ are equivalent as they correspond to a perturbation along the same physical surface. \label{fig3:Angle}}
\end{figure}

In the instability diagram, the phase plane is divided into several regions where gravitational instability (Equation~\ref{eq:nondim_grav}), inertial instability (Equation~\ref{eq:nondim_iner}), or mixed symmetric instability (Equation~\ref{eq:nondim-PVcriteria}) occur. In all regions except for the stable region $S_{S}$ located above the hyperbolic curve $xy=1$ in the first quadrant, at least one type of instability is present (see Figure~\ref{fig2:instability}c~\&~d and Figure~\ref{fig3:Angle}a).

The alignment of the most unstable mode with the buoyancy and angular momentum surfaces is indicative of different types of instabilities. To illustrate how the most unstable mode aligns with the buoyancy and angular momentum surfaces, we define $\alpha \in (-\pi/2,\pi/2]$ as the angle between the direction of the most unstable mode and the buoyancy surface, and $\beta \in (-\pi/2,\pi/2]$ as the angle between the direction of the most unstable mode and the angular momentum surface, with the most unstable mode counterclockwise of the buoyancy/angular momentum surface defined as positive (Figure~\ref{fig3:Angle}b). When $\alpha=0$, the most unstable mode aligns with the buoyancy surface, and there is no buoyancy transport associated with the most unstable mode (Equation~\ref{eq:transport_angle}). As a result, the most unstable mode can only extract kinetic energy from the background angular momentum field (see \ref{app:energy} for detailed derivations), and regions where $\alpha \approx 0$ can be identified as inertially unstable (Figure~\ref{fig3:Angle}c). Similarly, when the most unstable mode aligns with the angular momentum surface ($\beta = 0$), it can only extract potential energy from the background buoyancy field, and regions where $\beta \approx 0$ can be identified as gravitationally unstable (Figure~\ref{fig3:Angle}d). However, the opposite is not always true, and in many parts of the parameter regime, the most unstable mode does not align with either surface. The special case where the Rossby number is small and the most unstable mode always aligns with the angular momentum surface will be discussed in Section~\ref{subsec:lowRo}.

Most previous studies on symmetric instability have focused on what we here call the ``mixed symmetric instability'', thus neglecting regimes where $N^2<0$ or $\eta/f<0$, as they typically assume stable stratification in the gravitational direction and stable zonal velocity shear in the latitudinal direction. However, we find that $N^2 < 0$ is neither sufficient nor necessary for instability. In Figure~\ref{fig2:instability}c~\&~d, the dashed lines indicate $N^2=0$, with regions to the left of these lines corresponding to $N^2<0$. Therefore, when $b_r<0$, the system can remain stable even if $N^2<0$. Meanwhile, $\eta/f<0$ is a sufficient (but not necessary) condition for instability. When $\eta/f>0$, the sufficient and necessary condition for instability becomes $q_0 f_0 \sin{\theta_0} < 0$ (see detailed derivation in \ref{app:N2eta}). This result is consistent with \cite{flasar_turbulent_1978} and \cite{hathaway_convective_1979}, who assumed that the background flow is in thermal wind balance with no meridional shear, such that $\eta/f=1>0$. Under these assumptions, they found that symmetric instability would occur if and only if $q_0 f_0 \sin{\theta_0} < 0$ (equivalent to Equation~38 in \cite{hathaway_convective_1979} after substituting in the background thermal wind shear), consistent with our findings.

\subsection{Low Rossby number limit}\label{subsec:lowRo}

In the low Rossby number regime, $\mathrm{Ro} = U/(f_0 L) \ll 1$ where $U$ is the velocity scale and $L$ is the length scale of motion, the rotational effect dominates over nonlinear advection \citep[c.f. ][]{vallis2017atmospheric}. Many planetary flows in geophysical fluid dynamics likely fall into this regime, such as the deep convection in the bulk ocean away from the boundary layers on icy moons. For deep convection, the Rossby number can be estimated by the convective Rossby number, $\mathrm{Ro_C}=B^{1/2}f_0^{-3/2}H^{-1}$ \citep{maxworthy1994unsteady}, where $B$ is the buoyancy flux and $H$ is the depth of the ocean \citep[$\mathrm{Ro_C}$ is equivalent to the square root of the modified flux Rayleigh number in Rayleigh–Bénard convection studies, c.f.][]{christensen2006scaling}. In the bulk ocean interior, \cite{jansen2023energetic} suggest that the convective Rossby number is at most on the order $10^{-3}$, indicating that the deep convection in the bulk icy moon ocean is strongly constrained by rotation.

In the low Rossby number limit, we have $\overline{\omega}$, $r\partial \overline{\omega}/\partial r$, $r\partial \overline{\omega}/\partial z \ll f_0$, which implies $M_r \approx f_0 \approx 2\Omega$ and $M_z \ll f_0$. In gradient wind balance, we moreover have the scaling that $b_r, \ b_z \sim f_0 M_z \ll f_0^2$. Consequently, the criterion for inertial instability, Equation~\ref{eq:inertialvec}, reduces to 

\begin{equation}
    f_0^2 b_z^2 < 0,
\end{equation}

\noindent which cannot be satisfied, because small Rossby number flow is always inertially stable. Instability then occurs if, and only if, 

\begin{equation}
    b_z \sin{\theta_0}<0,
\end{equation}

\noindent i.e., for an unstable stratification in the direction parallel to planetary rotation, which coincides with the angular momentum surface in the small $\mathrm{Ro}$ limit. The most unstable mode (Equation~\ref{eq:most_unstable_mode}) is obtained when

\begin{equation}\label{eq:unstable_lowRo}
    \left| \tan{\delta_m} \right| \approx \left| \frac{-f_0^2 - [(f_0^2)^2]^{1/2}}{2 b_r \sin{\theta_0}} \right| = \left|\frac{f_0^2}{b_r \sin{\theta_0}} \right| \gg 1, \ \ \mathrm{i.e.,} \ \delta_m \approx \frac{\pi}{2},
\end{equation}

\noindent and the corresponding growth rate is

\begin{equation}\label{eq:grwoth_rate_lowRo}
    \sigma_{m} \approx (-b_z \sin{\theta_0})^{1/2}.
\end{equation}

Therefore, the most unstable mode is associated with slantwise convection aligned with the planetary angular momentum surface. This mode, parallel to the rotation axis, is likely to be important in the bulk oceans away from boundary layers on icy moons, which are thought to be characterized by low Rossby numbers. Such slantwise convection aligned with the rotation axis has been identified in numerical simulations of icy moon oceans \citep[e.g.,][]{ashkenazy2021dynamic,kang2022does,bire2022exploring,zeng2024effect}.

\section{Numerical simulations}\label{sec:numerical_simulations}

\subsection{Simulation set up}\label{subsec:simulation_setup}

\begin{figure}[b!]
\centering
\includegraphics[width=1.0\linewidth]{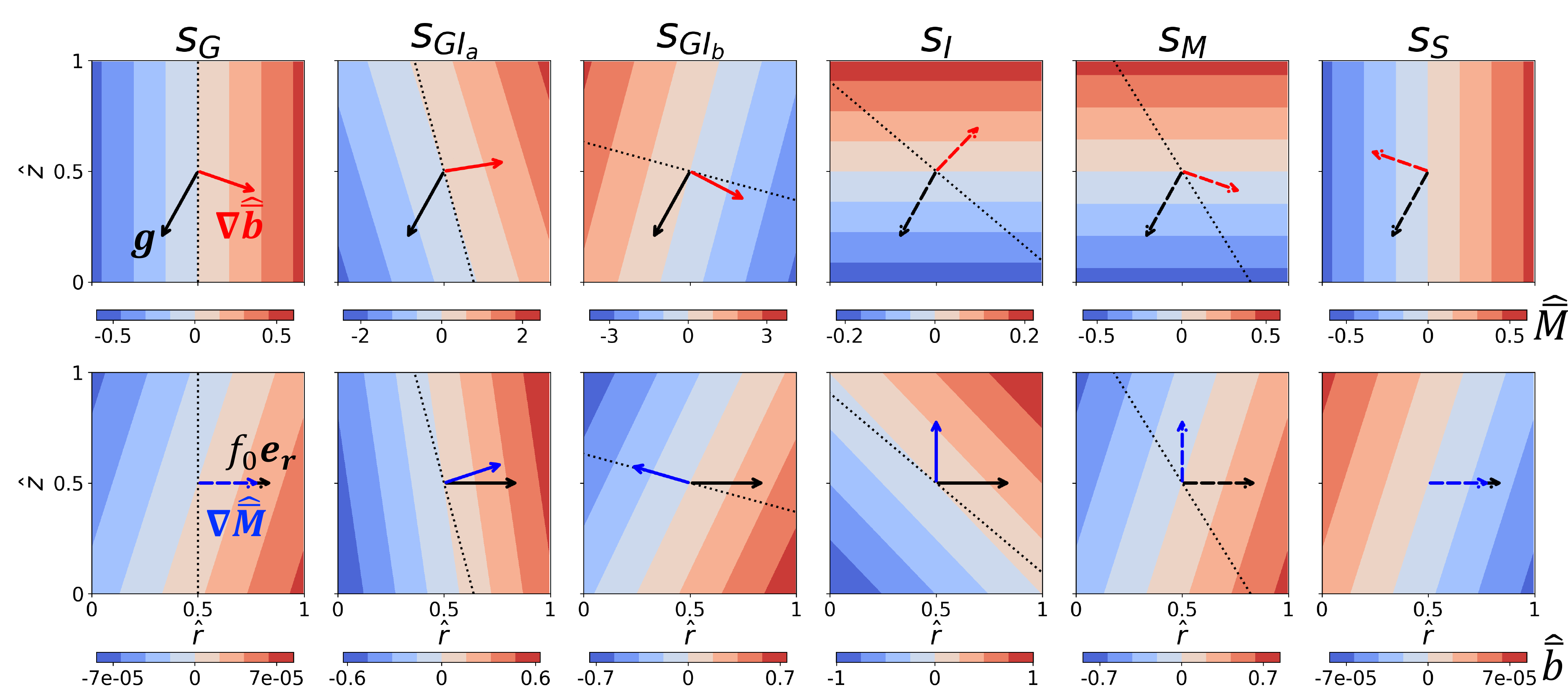}
\caption{Background states in the simulations. The first row shows the background angular momentum surfaces ($\hat{\overline{M}} = \hat{M}_r \hat{r} + \hat{M}_z \hat{z}$) and the second row shows the background buoyancy surfaces ($\hat{\overline{b}} = \hat{b}_r \hat{r} + \hat{b}_z \hat{z}$), where the domain-averaged values are subtracted. Blue arrows indicate the direction of angular momentum gradients, red arrows indicate the direction of buoyancy gradients, and black arrows indicate gravity $\boldsymbol{g}$ in the upper row and the Coriolis vector $f_0 \mathbf{e}_r$ in the lower row, all consistent with sketches in Figure~\ref{fig2:instability}a~\&~b. Solid arrows indicate gravitational (red) and/or inertial (blue) instability, while dashed arrows indicate stability. The black dotted line indicates the orientation of the most unstable mode predicted by Equation~\ref{eq:most_unstable_mode}.\label{fig4:Simulation}}
\end{figure}

To verify the theoretical results, we numerically integrate Equations \ref{eq4:u}--\ref{eq8:cont} for a zonally symmetric flow ($\partial/\partial \phi=0$) using Dedalus, which can solve initial-value partial differential equations using spectral methods \citep{burns2020dedalus}. We nondimensionalize the equations using the rotational time scale ($T=f_0^{-1}$) and the domain length scale. We neglect variations of the background field and the metric term in the continuity equation ($v/r$), allowing the use of a local Cartesian coordinate system and double-periodic boundary conditions (c.f. \ref{app:simulation}). We prescribe a background state which satisfies the gradient wind balance (Equations~\ref{eq9:thermalwind1}~\&~\ref{eq10:thermalwind2}), and solve for the evolution of the nondimensionalized perturbation fields ($\hat{u},\hat{v},\hat{w},\hat{b},\hat{\Phi}$). We apply a resolution with 256 grid points in both $r$ and $z$ directions. Simulations without viscosity and diffusivity develop unphysical grid-scale noise, once nonlinear effects become significant (Figure~\ref{figS2}). In addition to inviscid simulations, we therefore perform simulations with a Leith sub-grid parameterization \citep{leith1996stochastic} to represent the effects of sub-grid-scale eddy mixing, and compare their results with the simulations without viscosity and diffusivity. It takes some time for the most unstable mode to grow and become dominant, before which nonlinear effects may already induce turbulence in the system. To address this issue, we initialize the simulations with their most unstable modes (as inferred from an integration of the linearized equations). A detailed description of the simulation setup can be found in \ref{app:simulation}.

We carry out six simulations with the background field representing each region in the parameter space, where different types of instability (or no instability) occur (Figure~\ref{fig2:instability}c~\&~d). The simulation parameters are summarized in Table~\ref{tab:simulations} and the prescribed background fields are shown in Figure~\ref{fig4:Simulation}.

\begin{table}
  \begin{center}
\def~{\hphantom{0}}
  \begin{tabular}{ccccccccc}
      Simulation & $\hat{b}_r$ & $\hat{b}_z$ & $\hat{M}_r$ & $\hat{q}_0$ & $\hat{N}^2$ & Instability & $\hat{\sigma}_m$ \\
      \hline
      $s_{G}$ & 1e-4 & -3.33e-5 & 1 & -3.33e-5 & 2.11e-5 & Grav & 5.37e-3  \\
      $s_{GI_a}$ & 1 & 0.15 & 2.5 & -0.416 & 0.630 & Grav, Iner & 0.333  \\
      $s_{GI_b}$ & 1 & -0.5 & -4 & 0.884 & 6.70e-2 & Grav, Iner & 1.93 \\
      $s_{I}$ & 1 & 1 & 0 & -0.366 & 1.37 & Iner & 0.450 \\
      $s_{M}$ & 1 & -0.333 & 0 & -1.03 & 0.211 & Mixed & 0.920 \\
      $s_{S}$ & -1e-4 & 3.33e-5 & 1 & 3.33e-5 & -2.11e-5 & Stable & / \\
       &  & &  & &  & &  \\
  \end{tabular}
  \caption{Numerical simulation setups. The latitudes for all simulations are $\theta_0=60^\circ$. The simulation parameters are nondimensionalized with the rotational time scale ($f_0^{-1}$) and the domain length scale, and hats are used to indicate nondimensionalized quantities (See \ref{app:dispersion}). The types of instability in each simulation are indicated in the simulation name and the ``Instability'' column, where $G$ and Grav indicate gravitational instability, $I$ and Iner indicate inertial instability, $M$ and Mixed indicate mixed symmetric instability, and $S$ and Stable indicate the system is stable. Simulations $s_{G}$ and $s_{S}$ represent the low Rossby number limit.}
  \label{tab:simulations}
  \end{center}
\end{table}

\begin{figure}[b!]
\centering
\includegraphics[width=1.0\linewidth]{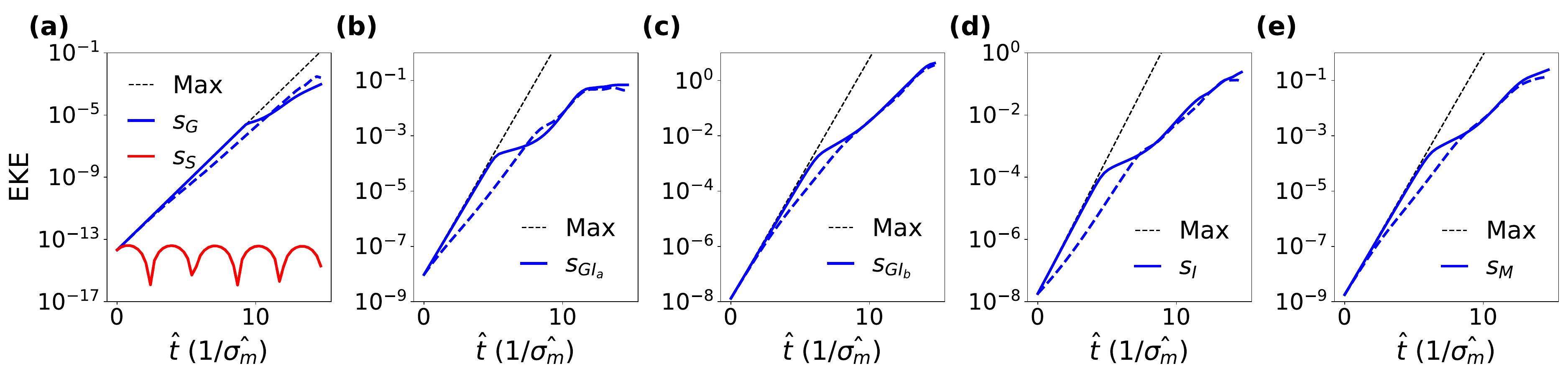}
\caption{Time series for domain-averaged eddy kinetic energy (EKE), $(\hat{u}^2$+$\hat{v}^2$+$\hat{w}^2$)/2. The blue and red solid lines show the simulation results without diffusivity and viscosity, and the blue and red dashed lines show the simulation results with eddy diffusivity and viscosity. Note that the red dashed line overlaps the red solid line because the effect of eddy viscosity is weak. The black dashed line shows the predicted maximum growth rate of EKE ($2\hat{\sigma}_m$ due to the square, Equation~\ref{eq:max_growth_rate}). The time is normalized by the maximum growth rate in each simulation (for $s_{S}$, the normalization uses the maximum growth rate for simulation $s_{G}$). \label{fig4:timeseries_nonlinear}}
\end{figure}

\begin{figure}
\centering
\includegraphics[width=1.0\linewidth]{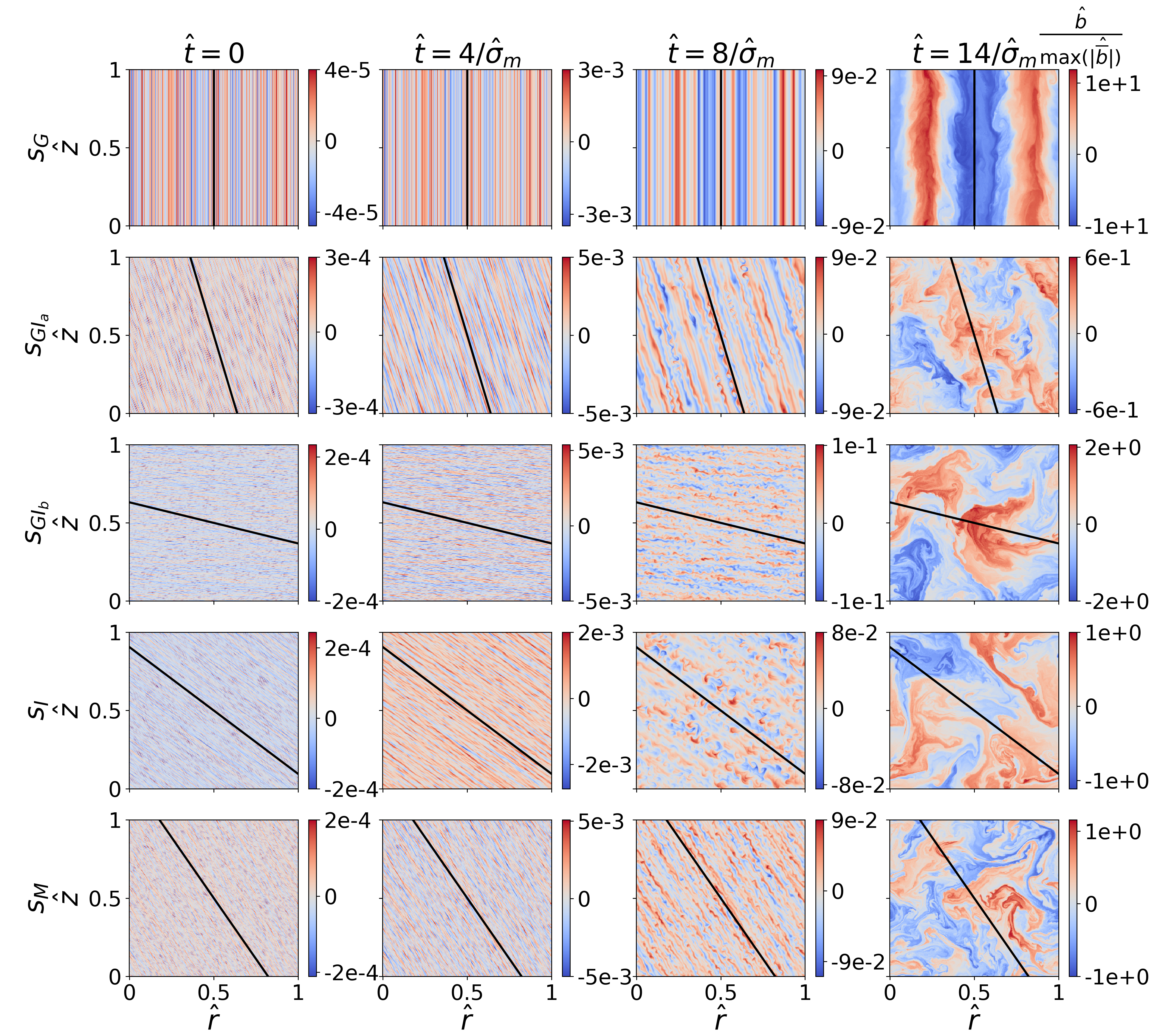}
\caption{Snapshots of the normalized buoyancy anomaly fields from simulations with eddy diffusivity and viscosity. The buoyancy anomaly is normalized by the maximum value of the background buoyancy field ($\hat{b}/\mathrm{max}(|\hat{\overline{b}}|)$) in each simulation. The time of each snapshot is indicated above the respective columns and is normalized by the maximum growth rate $\hat{\sigma}_{m}$ in each simulation. The black solid lines indicate the orientation of the most unstable mode predicted by Equation~\ref{eq:most_unstable_mode}. \label{fig5:snapshots}}
\end{figure}

\subsection{Simulation results}\label{subsec:simulation_results}

For all simulations without diffusivity and viscosity, the eddy kinetic energy (EKE) grows exponentially, consistent with the maximum growth rate predicted by Equation~\ref{eq:max_growth_rate}. When the Leith closure is applied, the EKE growth rate decreases slightly due to viscous and diffusive dissipation (Figure~\ref{fig4:timeseries_nonlinear}). During the exponential growth phase ($\hat{t}=4/\hat{\sigma}_m$) of simulations with an unstable background state, the orientation of the perturbation fields aligns closely with the most unstable mode predicted by Equation~\ref{eq:most_unstable_mode} (Figure~\ref{fig5:snapshots}, second column). At $\hat{t}=8/\hat{\sigma}_m$, nonlinear effects begin to become important, and the flow field becomes turbulent, while it is still dominated by small-scale eddies roughly aligned with the most unstable mode (Figure~\ref{fig5:snapshots}, third column). Eventually, at $\hat{t}=14/\hat{\sigma}_m$, the non-linear advection effects become dominant, the flow is characterized by domain-scale eddies, and the growth rates of EKE decrease significantly (Figure~\ref{fig5:snapshots}, last column). Notably, the EKE eventually converges for simulations with and without eddy viscosity and diffusivity (Figure~\ref{fig4:timeseries_nonlinear}).

The simulation with stable background state ($s_{S}$) does not show exponential growth but an oscillation behavior. The initial condition used in simulation $s_{S}$ is the most unstable mode in simulation $s_{G}$, which has $\delta = \pi/2$. In this case, we can estimate the oscillation period $P$ according to the dispersion relation (Equation~\ref{eq18:dispersion}) as $P=584.79$, or 3.14/$\sigma_m$ with $\sigma_m$ being the maximum growth rate of $s_{G}$, consistent with the oscillation period in the simulation (Figure~\ref{fig4:timeseries_nonlinear}a).

The numerical simulations also show that $N^2<0$ is neither a sufficient nor necessary condition for instability. In simulation $s_{S}$, $N^2<0$, but the system is stable. In simulations $s_{G}$ and $s_{GI_b}$, $N^2>0$, but the systems are unstable, with either $q_0 f_0 \sin{\theta_0}>0$ ($s_{GI_b}$) or $q_0 f_0 \sin{\theta_0}<0$ ($s_{G}$).

In simulations with $\mathrm{Ro} \ll 1$ ($s_{G}$ and $s_{S}$), the instability criteria reduce to $b_z \sin{\theta_0} < 0$. In the unstable case (simulation $s_{G}$), the most unstable mode is parallel to the rotation axis, i.e., slantwise convection parallel to planetary rotation. 

\section{Discussions}\label{sec:discussions}

The analysis conducted in this manuscript provides physical insight into symmetric instability within a generalized framework, applicable across a wide range of planetary parameter regimes. This framework offers useful guidance for developing more accurate convection schemes for icy moon oceans.

The traditional convective adjustment schemes widely applied in ocean General Circulation Models \citep[e.g.,][]{marotzke1991influence} parameterize upright convection when $N^2 < 0$. However, our results show that a negative stratification in the gravitational direction ($N^2 < 0$) is neither a sufficient nor necessary condition for instability. This suggests that such schemes may not be appropriate for capturing slantwise convection in icy moon oceans. Given that $b_z \sin{\theta_0} < 0$ is a sufficient condition for symmetric instability, and becomes a necessary and sufficient condition in the low Rossby number limit, we propose that stratification along the rotation axis should be treated as the criterion for static stability and used to trigger convective parameterizations in planetary ocean models.

In the linear regime, the angle of the most unstable mode controls both the direction and relative magnitude of the heat and momentum transports (see Section~\ref{subsec:physical_stability} and \ref{app:energy}). However, our linear instability analysis does not capture nonlinear effects. As shown in Figure~\ref{fig5:snapshots}, once the flow becomes fully nonlinear and turbulent, it no longer follows the structure of the most unstable mode. A complete understanding of heat and momentum transport in fully nonlinear slantwise convection is left for future studies.

Although we consider Boussinesq fluids in this manuscript, the results are also relevant for anelastic fluids, such as atmospheric convection on gas giants \citep[e.g.,][]{oneill_slantwise_2016}. In a compressible fluid, the Solberg–Høiland criteria in Equations~\ref{eq:trace} and \ref{eq:det} remain valid, but the buoyancy gradients must be replaced by the specific entropy gradient, as derived by \citet{ogilvie2019internal}. In the low-Rossby number regime, where the angular momentum gradient is dominated by planetary rotation, the criteria reduce to unstable entropy gradients along the rotation axis, similar to our findings under the Boussinesq approximation. Observations of the gravitational fields of Jupiter and Saturn suggest that the zonal jets on these planets are aligned with the rotation axis \citep{kaspi2018jupiter,galanti2019saturn,kaspi2023observational}, indicating that the prevailing modes are parallel to planetary rotation. The in-situ temperature profile measurement from the Galileo probe indicates a neutral stratification along constant angular momentum surfaces, which are parallel to the rotation axis outside the high-shear region in the upper weather layer \citep{o2017galileo}. Numerical simulations of gas giant atmospheres also reveal slantwise convection aligned with planetary rotation \citep{christensen2002zonal,kaspi2009deep,heimpel2016simulation}.

Our study focuses exclusively on zonally symmetric instabilities. However, asymmetric modes could also play a significant role in fluid motions on a rotating planet. For instance, equatorial convective rolls driven by heating from the bottom of the fluid are believed to be important in driving the equatorial jets on gas giants \citep[e.g.,][]{busse1982differential,busse1986convection}. Therefore, it is important to characterize the growth rate of different symmetric and asymmetric unstable modes to determine under which regime the symmetric mode dominates. \cite{stone1966non} studied the growth rate of symmetric instability, Kelvin-Helmholtz instability, and baroclinic instability on an $f$-plane with hydrostatic balance. He concludes that on an $f$-plane, symmetric instability prevails when $0.25<\mathrm{Ri}<0.95$; Kelvin-Helmholtz instability dominates for $\mathrm{Ri}<0.25$; and baroclinic instability is dominant when $\mathrm{Ri}>0.95$. When considering the horizontal component of planetary rotation, \cite{jeffery_effect_2009} suggest that the transition point between symmetric and baroclinic instability dominance can exceed $\mathrm{Ri}=0.95$. Understanding how the symmetric mode, such as slantwise convection in the low Rossby number limit, interacts with baroclinic instability under arbitrary background stratification and shear, and conditions that separate symmetric and asymmetric mode dominance, are important future research directions. We applied a linear instability analysis method, which is valuable to understand stability criteria for small-amplitude perturbations but is not applicable to large-amplitude perturbations or non-linear instability problems. \cite{bowman1995nonlinear} applied the energy-Casimir stability method \citep{holm1985nonlinear,shepherd1990symmetries,cho1993application} to study nonlinear symmetric instability where the hydrostatic approximation is made and the rotation is aligned with gravity. However, extending this approach to a more generalized setup \citep[e.g.,][]{fruman_symmetric_2008} presents challenges that remain to be addressed in future work. In our current study, solutions are sought on an infinitely large domain (local plane-wave solutions). Future work should also explore how boundary conditions might modify the characteristics of the most unstable modes.

\section{Conclusions}\label{sec:conclusions}

We conduct local linear instability analysis with slowly-varying background shear and stratification, and find that the necessary and sufficient conditions for instability of zonally symmetric Boussinesq flow on a rotating planet are the background field is (1) gravitationally unstable, or (2) inertially unstable, or (3) the background potential vorticity has a different sign from the planetary vorticity, which is consistent with previous studies on symmetric instability \citep[e.g.,][]{hoskins_role_1974}. In our framework, we define the gravitational instability criterion as unstable buoyancy stratification along angular momentum surfaces (Equation~\ref{eq:gravitationalvec}~\&~Figure~\ref{fig2:instability}a) and the inertial instability criterion as unstable angular momentum shear along buoyancy surfaces (Equation~\ref{eq:inertialvec}~\&~Figure~\ref{fig2:instability}b), which ensures both to be sufficient conditions for instability. Mathematically, the criterion for instability can most compactly be expressed as either $b_z \sin{\theta_0}<0$ (i.e., unstable stratification along the planetary rotation axis) or $q_0 f_0 \sin{\theta_0}<0$ where $q_0$ is the background potential vorticity. When instability occurs, the growth rate is not sensitive to the magnitude of the radial and vertical wavenumbers, $k_r$ and $k_z$, but is only a function of their ratio, $k_r/k_z$, i.e., the tilting direction of the mode. 

A negative stratification in the gravitational direction ($N^2<0$) is neither necessary nor sufficient for instability, indicating that traditional convective adjustment schemes used in ocean general circulation models \citep[e.g.,][]{marotzke1991influence} may not be appropriate for representing slantwise convection in icy moon oceans.

In the low Rossby number limit, instability occurs if and only if $b_z \sin{\theta_0} < 0$, and the most unstable mode is slantwise convection parallel to the planetary rotation axis. The low Rossby number limit proves to be particularly valuable for understanding slantwise convection in icy moon oceans, as this phenomenon is neither properly resolved nor parameterized in global ocean simulations \citep[c.f. ][]{zeng2024effect}. 

Our work provides criteria and physical insight into instabilities within a generalized framework applicable across diverse planetary parameter regimes. These results enhance our understanding of fluid motions on various planetary bodies and can improve parameterizations of sub-grid-scale transports in large-scale models, such as slantwise convection in icy moon oceans.

\appendix

\section{Derivation of the dispersion relation}\label{app:dispersion}

After subtracting the gradient wind balance and the hydrostatic balance for the background flow (Equations~\ref{eq9:thermalwind1}~\&~\ref{eq10:thermalwind2}) from the original Equations~\ref{eq4:u}-\ref{eq8:cont}, we have the perturbation equations:

\begin{equation}\label{eq11:up}
     \left( \frac{\partial}{\partial t} + v'\frac{\partial}{\partial r} + w'\frac{\partial}{\partial z} \right) u' + \left(2\Omega + \frac{\partial \overline{u}}{\partial r} + \frac{\overline{u}}{r} \right) v' + \pz{\overline{u}}w' = 0,
\end{equation}

\begin{equation}\label{eq12:vp}
    \left( \frac{\partial}{\partial t} + v'\frac{\partial}{\partial r} + w'\frac{\partial}{\partial z} \right) v' - \left(2\Omega + \frac{2\overline{u}}{r}\right) u' - \cos{\theta}b' + \pr{\Phi'} = 0,
\end{equation}

\begin{equation}\label{eq13:wp}
    \left( \frac{\partial}{\partial t} + v'\frac{\partial}{\partial r} + w'\frac{\partial}{\partial z} \right) w' - \sin{\theta} b' + \pz{\Phi'} = 0,
\end{equation}

\begin{equation}\label{eq14:bp}
    \left( \frac{\partial}{\partial t} + v'\frac{\partial}{\partial r} + w'\frac{\partial}{\partial z} \right) b' + \pr{\overline{b}} v' + \pz{\overline{b}} w' = 0,
\end{equation}

\begin{equation}\label{eq15:contp}
    \pr{v'} + \pz{w'} + \frac{v'}{r} = 0.
\end{equation}

We assume that the background fields vary slowly over the characteristic length scale of the perturbations, $L$, so that they can be treated as constants. This criterion can be formally obtained by performing the Taylor expansion of all background fields locally around $(r = r_0, z = z_0)$:

\begin{equation}\label{eq:f-expansion}
\begin{aligned}
    2\Omega + \frac{2\overline{u}}{r} &= 2\Omega + 2\overline{\omega} \\
    &= 2\Omega + 2\overline{\omega}(r_0,z_0) + 2 \left.\frac{\partial \overline{\omega}}{\partial r}\right|_{(r_0,z_0)} r' + 2 \left.\frac{\partial \overline{\omega}}{\partial z}\right|_{(r_0,z_0)} z' + ...,
\end{aligned}
\end{equation}

\begin{equation}\label{eq:Mr-expansion}
\begin{aligned}
    2\Omega + \frac{\partial \overline{u}}{\partial r} + \frac{\overline{u}}{r} &= 2\Omega + 2\overline{\omega} + r\pr{\overline{\omega}} \\
    &= 2\Omega + 2\overline{\omega}(r_0,z_0) +r_0\left.\pr{\overline{\omega}}\right|_{(r_0,z_0)} \\
    & \ \ \ + 
    \left(3 \frac{\partial \overline{\omega}}{\partial r} + r_0 \frac{\partial^2 \overline{\omega}}{\partial r^2}\right)_{(r_0,z_0)} r' + \left(2\pz{\overline{\omega}} + r_0 \frac{\partial^2 \overline{\omega}}{\partial r\partial z} \right)_{(r_0,z_0)} z' + ...,
\end{aligned}
\end{equation}

\begin{equation}\label{eq:Mz-expansion}
    \pz{\overline{u}} = r\pz{\overline{\omega}} = r_0\left.\pz{\overline{\omega}}\right|_{(r_0,z_0)} + \left(r_0\frac{\partial^2 \overline{\omega}}{\partial r \partial z} + \frac{\partial \overline{\omega}}{\partial z}\right)_{(r_0,z_0)}r' + r_0\left.\frac{\partial^2 \overline{\omega}}{\partial z^2}\right|_{(r_0,z_0)}z' + ...,
\end{equation}

\begin{equation}\label{eq:br-expansion}
    \pr{\overline{b}} = \left.\pr{\overline{b}}\right|_{(r_0,z_0)} + \left.\frac{\partial^2 \overline{b}}{\partial r^2}\right|_{(r_0,z_0)}r' + \left.\frac{\partial^2 \overline{b}}{\partial r \partial z}\right|_{(r_0,z_0)}z'  + ...,
\end{equation}

\begin{equation}\label{eq:bz-expansion}
    \pz{\overline{b}} = \left.\pz{\overline{b}}\right|_{(r_0,z_0)} + \left.\frac{\partial^2 \overline{b}}{\partial r \partial z}\right|_{(r_0,z_0)}r' + \left.\frac{\partial^2 \overline{b}}{\partial z^2}\right|_{(r_0,z_0)}z' + ...,
\end{equation}

\begin{equation}\label{eq:sintheta-expansion}
\begin{aligned}
    \sin{\theta} &= \sin{\theta_0} + \left.\pr{\sin{\theta}}\right|_{(r_0,z_0)}r' + \left.\pz{\sin{\theta}}\right|_{(r_0,z_0)}z' + ... \\
    &= \sin{\theta_0} - \sin{\theta_0} \cos{\theta_0} \frac{r'}{\sqrt{r_0^2+z_0^2}} + \cos^2{\theta_0}  \frac{z'}{\sqrt{r_0^2+z_0^2}}  + ...,
\end{aligned}
\end{equation}

\begin{equation}\label{eq:costheta-expansion}
\begin{aligned}
    \cos{\theta} &= \cos{\theta_0} + \left.\pr{\cos{\theta}}\right|_{(r_0,z_0)}r' + \left.\pz{\cos{\theta}}\right|_{(r_0,z_0)}z' + ... \\
    &= \cos{\theta_0} + \sin^2{\theta_0} \frac{r'}{\sqrt{r_0^2+z_0^2}} - \sin{\theta_0} \cos{\theta_0}  \frac{z'}{\sqrt{r_0^2+z_0^2}}  + ...,
\end{aligned}
\end{equation}

\noindent where $\theta_0 = \arctan{(z_0 / r_0)}$ is the local latitude, and $|r'| \sim |z'| \sim L$. We study the local instability of fluid in the outer region of the planet, where we assume $L \ll r_0$. The left-hand-sides of Equations~\ref{eq:f-expansion}--\ref{eq:costheta-expansion} remain approximately constant over the scale $L$ if

\begin{equation}\label{eq:background_approx_condition}
\begin{aligned}
    &\left( \frac{|\overline{\omega}|}{|\mathrm{d} \overline{\omega}|} \right)_{(r_0,z_0)} \gg L, \ \ \left( \frac{|\mathrm{d} \overline{\omega}|}{|\mathrm{d}^2 \overline{\omega}|} \right)_{(r_0,z_0)} \gg L, \\ 
    &\left( \frac{|\mathrm{d} \overline{b}|}{|\mathrm{d}^2 \overline{b}|} \right)_{(r_0,z_0)} \gg L, \ \ \frac{|\cos{\theta_0}|}{|\sin^2{\theta_0}|}, \frac{|\sin{\theta_0}|}{|\cos^2{\theta_0}|}  \gg \frac{L}{r_0},
\end{aligned}
\end{equation}

\noindent where $\mathrm{d}$ and $\mathrm{d}^2$ represent the first and second spatial derivatives, respectively. Equation~\ref{eq:background_approx_condition} indicates that the approximation holds when $\overline{b}$ and $\overline{\omega}$ have a large radius of curvature relative to the characteristic perturbation length scale $L$, the variation in $\overline{\omega}$ is small compared to its average, and the region we consider is not too close to either the equator or the pole. Under this limit, we define

\begin{equation}\label{eq:all_field_constant}
\begin{aligned}
    f_0 \equiv 2\Omega + 2\overline{\omega}(r_0,z_0), \quad &M_r \equiv f_0 + r_0\left.\frac{\partial \overline{\omega}}{\partial r}\right|_{(r_0,z_0)}, \quad M_z \equiv r_0\left.\frac{\partial \overline{\omega}}{\partial z}\right|_{(r_0,z_0)}, \\
    &b_r \equiv \left.\frac{\partial \overline{b}}{\partial r}\right|_{(r_0,z_0)}, \quad b_z \equiv \left.\frac{\partial \overline{b}}{\partial z}\right|_{(r_0,z_0)}.
\end{aligned}
\end{equation}

We nondimensionalize Equations~\ref{eq11:up}-\ref{eq15:contp} using the rotational time scale ($f_0^{-1}$) and the perturbation length scale ($L$), such that 
\begin{equation}
\begin{aligned}
    \hat{t} = tf_0, \ (\hat{u},\hat{v},\hat{w})=(u',v',w')/(f_0 L), \ \hat{b} = b'/(f_0^2 L), \ \hat{\Phi} = \Phi'/(f_0^2L^2), \\
    (\hat{r},\hat{z}) = (r',z')/L, \ (\hat{M}_r,\hat{M}_z) = (M_r,M_z)/f_0, \ (\hat{b}_r,\hat{b}_z) = (b_r,b_z)/f_0^2.
\end{aligned}
\end{equation}

\noindent The nondimensionalized equations become

\begin{equation}\label{eqa1:u}
     \left( \frac{\partial}{\partial \hat{t}} + \hat{v}\frac{\partial}{\partial \hat{r}} + \hat{w}\frac{\partial}{\partial \hat{z}} \right) \hat{u} + \hat{M}_r \hat{v} + \hat{M}_z \hat{w} = 0,
\end{equation}

\begin{equation}\label{eqa2:v}
    \left( \frac{\partial}{\partial \hat{t}} + \hat{v}\frac{\partial}{\partial \hat{r}} + \hat{w}\frac{\partial}{\partial \hat{z}} \right) \hat{v} - \hat{u} - \cos{\theta_0} \hat{b} + \frac{\partial \hat{\Phi}}{\partial \hat{r} } = 0,
\end{equation}

\begin{equation}\label{eqa3:w}
    \left( \frac{\partial}{\partial \hat{t}} + \hat{v}\frac{\partial}{\partial \hat{r}} + \hat{w}\frac{\partial}{\partial \hat{z}} \right) \hat{w} - \sin{\theta_0} \hat{b} + \frac{\partial \hat{\Phi}}{\partial \hat{z} } = 0,
\end{equation}

\begin{equation}\label{eqa4:b}
    \left( \frac{\partial}{\partial \hat{t}} + \hat{v}\frac{\partial}{\partial \hat{r}} + \hat{w}\frac{\partial}{\partial \hat{z}} \right) \hat{b} + \hat{b}_r \hat{v} + \hat{b}_z \hat{w} = 0,
\end{equation}

\begin{equation}\label{eqa5:cont}
    \frac{\partial \hat{v} }{\partial \hat{r} } + \frac{\partial \hat{w} }{\partial \hat{z} } + \frac{\hat{v}}{r_0/L + \hat{r}} = 0.
\end{equation}

We neglect the last term in Equation~\ref{eqa5:cont} because $r_0/L \gg 1$ for the local instability problem. In the linearized instability problem, we assume that the perturbation fields are much smaller than the background fields, and obtain the first-order linear perturbation equations:

\begin{equation}\label{eqa6:ulinear}
     \frac{\partial \hat{u}}{\partial \hat{t}} + \hat{M}_r \hat{v} + \hat{M}_z \hat{w} = 0,
\end{equation}

\begin{equation}\label{eqa7:vlinear}
    \frac{\partial \hat{v}}{\partial \hat{t}} - \hat{u} - \cos{\theta_0} \hat{b} + \frac{\partial \hat{\Phi}}{\partial \hat{r} } = 0,
\end{equation}

\begin{equation}\label{eqa8:wlinear}
    \frac{\partial \hat{w}}{\partial \hat{t}} - \sin{\theta_0} \hat{b} + \frac{\partial \hat{\Phi}}{\partial \hat{z} } = 0,
\end{equation}

\begin{equation}\label{eqa9:blinear}
    \frac{\partial \hat{b}}{\partial \hat{t}} + \hat{b}_r \hat{v} + \hat{b}_z \hat{w} = 0,
\end{equation}

\begin{equation}\label{eqa10:contlinear}
    \frac{\partial \hat{v} }{\partial \hat{r} } + \frac{\partial \hat{w} }{\partial \hat{z} } = 0.
\end{equation}

We look for plane-wave solutions of the linearized perturbation equations~\ref{eqa6:ulinear}-\ref{eqa10:contlinear}: 

\begin{equation}\label{eq:planewave}
    (\hat{u}, \hat{v}, \hat{w}, \hat{b}, \hat{\Phi}) = (A_u, A_v, A_w, A_b, A_\Phi)\exp(i\hat{k}_r \hat{r} + i\hat{k}_z \hat{z} - i\hat{\fre} \hat{t}),
\end{equation}

\noindent where $A_u, A_v, A_w, A_b, A_\Phi$ are constant amplitudes, and $(\hat{k}_r,\hat{k}_z)=(k_r,k_z)L$.

Under these conditions and substituting Equation~\ref{eq:planewave} into Equations~\ref{eqa1:u}--\ref{eqa5:cont}, we get a linearized equation system. The linearized equations must have zero determinant to have non-trivial solutions, which requires

\begin{equation}
    \left| \begin{matrix}
        -i\hat{\fre} & \hat{M}_r & \hat{M}_z & 0 & 0 \\
        -1 & -i\hat{\fre} & 0 & -\cos{\theta_0} & i \hat{k}_r \\
        0 & 0 & -i\hat{\fre} & -\sin{\theta_0} & i \hat{k}_z \\
        0 & \hat{b}_r & \hat{b}_z & -i\hat{\fre} & 0 \\
        0 & i\hat{k}_r & i\hat{k}_z & 0 & 0 \\
    \end{matrix}\right| = 0.
\end{equation}

This gives

\begin{equation}\label{eq17:omega}
    \hat{\fre} [(\hat{k}_r^2+\hat{k}_z^2)\hat{\fre}^2 - \hat{b}_z \sin{\theta_0} \hat{k}_r^2 - (\hat{M}_r + \hat{b}_r \cos{\theta_0}) \hat{k}_z^2 + 2 \hat{b}_r \sin{\theta_0} \hat{k}_r \hat{k}_z ] = 0.
\end{equation}

Neglecting the trivial solutions $\hat{\fre} = 0$ and $\hat{k}_r = \hat{k}_z = 0$, we have the dispersion relation:

\begin{equation}\label{eq:nondim-dispersion}
    \hat{\fre}^2 = \frac{\hat{b}_z \sin{\theta_0} \hat{k}_r^2 + (\hat{M}_r + \hat{b}_r \cos{\theta_0}) \hat{k}_z^2 - 2 \hat{b}_r \sin{\theta_0} \hat{k}_r \hat{k}_z}{\hat{k}_r^2 + \hat{k}_z^2}.
\end{equation}

Restoring Equation~\ref{eq:nondim-dispersion} back to dimensionalized form, we have the dispersion relation:

\begin{equation}
    \fre^2 = \frac{b_z \sin{\theta_0} k_r^2 + (f_0 M_r + b_r \cos{\theta_0}) k_z^2 - 2 b_r \sin{\theta_0} k_r k_z}{k_r^2 + k_z^2}.
\end{equation}

\section{Heat and zonal momentum transport by the most unstable mode}\label{app:energy}

In this section, we analyze the heat transport ($\mathrm{Re}(\hat{v}\hat{b}^*)$, $\mathrm{Re}(\hat{w}\hat{b}^*)$) and zonal momentum transport ($\mathrm{Re}(\hat{v}\hat{u}^*)$, $\mathrm{Re}(\hat{w}\hat{u}^*)$) associated with the most unstable mode, where $\mathrm{Re}$ indicates the real part and the asterisk indicates the complex conjugate. By substituting Equation~\ref{eq:planewave} into Equations~\ref{eqa6:ulinear}, \ref{eqa9:blinear}, and \ref{eqa10:contlinear}, we have

\begin{equation}\label{eq:momentum_heat}
\begin{aligned}
    \hat{w}\hat{u}^* = \frac{\hat{M}_r\hat{k}_z-\hat{M}_z \hat{k}_r}{-i\hat{\fre} \hat{k}_r} |\hat{w}|^2, \ \hat{v}\hat{u}^* = \frac{\hat{k}_z(\hat{M}_r\hat{k}_z-\hat{M}_z \hat{k}_r)}{i\hat{\fre} \hat{k}_r^2} |\hat{w}|^2, \\
    \hat{w}\hat{b}^* = \frac{\hat{b}_r\hat{k}_z-\hat{b}_z \hat{k}_r}{-i\hat{\fre} \hat{k}_r} |\hat{w}|^2, \ \hat{v}\hat{b}^* = \frac{\hat{k}_z(\hat{b}_r\hat{k}_z-\hat{b}_z \hat{k}_r)}{i\hat{\fre} \hat{k}_r^2} |\hat{w}|^2.
\end{aligned}
\end{equation}

For stable modes, the frequency $\hat{\fre}$ is real, so that $\hat{v}$ and $\hat{w}$ are $\pi/2$ out of phase with $\hat{u}$ and $\hat{b}$. As a result, we have $\mathrm{Re}(\hat{w}\hat{u}^*) = \mathrm{Re}(\hat{v}\hat{u}^*) = \mathrm{Re}(\hat{w}\hat{b}^*) = \mathrm{Re}(\hat{v}\hat{b}^*) = 0$, implying zero net heat and momentum transport.

For unstable modes, $-i\hat{\fre}$ is replaced by the positive growth rate $\hat{\sigma}$, which yields

\begin{equation}\label{eq:transport_angle}
    \frac{\mathrm{Re}(\hat{w}\hat{u}^*)}{\mathrm{Re}(\hat{v}\hat{u}^*)} = \frac{\mathrm{Re}(\hat{w}\hat{b}^*)}{\mathrm{Re}(\hat{v}\hat{b}^*)} = -\frac{\hat{k}_r}{\hat{k}_z} = \tan{\delta},
\end{equation}

\noindent showing that the net heat and zonal momentum transports associated with the unstable mode are always aligned with the mode’s orientation.

Equation~\ref{eq:momentum_heat} also provides the relative magnitude of the heat and zonal momentum transport:

\begin{equation}\label{eq:relative_heat_momentum}
    \frac{|\mathbf{u}u|}{|\mathbf{u}b|}= \frac{\hat{M}_r\hat{k}_z-\hat{M}_z \hat{k}_r}{\hat{b}_r\hat{k}_z-\hat{b}_z \hat{k}_r} = \frac{|\nabla \hat{\overline{M}}| \sin{\beta}}{|\nabla \hat{\overline{b}}| \sin{\alpha}},
\end{equation}

\noindent where $\mathbf{u}=v\mathbf{e_r}+w\mathbf{e_z}$, and $\alpha$ and $\beta$ are the angles between the most unstable mode and the buoyancy and angular momentum surfaces, respectively (see Figure~\ref{fig3:Angle}b). In addition to the overall magnitudes of the buoyancy and angular momentum gradients, the angles $\alpha$ and $\beta$ therefore determine the relative magnitude of the buoyancy and momentum fluxes. When the most unstable mode aligns with the angular momentum surface ($\beta=0$), there is no zonal momentum transport, and the mode cannot extract kinetic energy from the background angular momentum field. Similarly, when $\alpha=0$, the most unstable mode aligns with the buoyancy surface, and the mode cannot extract potential energy from the background buoyancy field.

\section{Instability analysis for $b_r \sin{\theta_0} = 0$}\label{app:brsin}

In this section, we discuss the special case when $b_r \sin{\theta_0} = 0$, which essentially suggests $b_r=0$ since $\sin{\theta_0}=0$ violates Equation~\ref{eq:background_approx_condition}. In this case, the dispersion relation (Equation~\ref{eq18:dispersion}) becomes

\begin{equation}
    \fre^2 = \frac{b_z \sin{\theta_0} k_r^2 + f_0 M_r k_z^2}{k_r^2 + k_z^2}.
\end{equation}

Therefore, the instability criteria reduce to $b_z \sin{\theta_0}<0$ or $f_0 M_r<0$, i.e., the instability for perturbations in the vertical or radial directions (where the vertical is here defined as parallel to the rotation axis, while the radial is orthogonal to the rotation axis). When the system is unstable, the most unstable mode can be inferred by taking the limit $b_r \to 0$ of Equation~\ref{eq:most_unstable_mode}, which gives 

\begin{equation}
    \lim_{b_r \to 0} \tan{\delta_m} \to 
    \begin{cases}
        -\frac{b_r \sin{\theta_0}}{b_z \sin{\theta_0} - f_0 M_r - b_r \cos{\theta_0}} \to 0 & \mathrm{if} \ b_z \sin{\theta_0} - f_0 M_r >0, \\
        \frac{b_z \sin{\theta_0} - f_0 M_r - b_r \cos{\theta_0}}{b_r \sin{\theta_0}} \to \pm \infty & \mathrm{if} \ b_z \sin{\theta_0} - f_0 M_r <0,
    \end{cases}
\end{equation}

\noindent with the maximum growth rate being

\begin{equation}
    \lim_{b_r \to 0} \sigma_m \to 
    \begin{cases}
        (- f_0 M_r)^{1/2} & \mathrm{if} \ b_z \sin{\theta} - f_0 M_r >0, \\
        (- b_z \sin{\theta_0})^{1/2} & \mathrm{if} \ b_z \sin{\theta} - f_0 M_r  <0.
    \end{cases}
\end{equation}

This indicates that the most unstable mode always aligns with either the radial ($\delta_m = 0$) or vertical ($\delta_m = \pi/2$) directions, depending on the relative magnitude of the growth rate in these two directions, $(- f_0 M_r )^{1/2}$ and $(-b_z \sin{\theta_0})^{1/2}$.

\section{Instability analysis for $N^2<0$ and $\eta/f<0$}\label{app:N2eta}

In this section, we show that $N^2<0$ is not a sufficient condition for instability in a zonally symmetric flow, while $\eta/f<0$ is a sufficient condition for instability.

$N^2$ can be expressed as

\begin{eqnarray}
N^2 &=& \frac{\partial \overline{b}}{\partial R} \nonumber \\
&=& b_z \sin{\theta_0} + b_r \cos{\theta_0} \nonumber \\
&=& |b_r \sin{\theta_0}| \left[\frac{b_z \sin{\theta_0}}{|b_r \sin{\theta_0}|} + \frac{b_r}{|b_r|} \frac{\cos{\theta_0}}{|\sin{\theta_0}|} \right] \nonumber \\
&=& |b_r \sin{\theta_0}| [x + |\cot{\theta_0}| \mathrm{sgn}(b_r)],
\end{eqnarray}

\noindent where $x = b_z \sin{\theta_0} / |b_r \sin{\theta_0}|$ is defined in Equation~\ref{eq:definexy}, and we used the definition that $\theta_0 \in (-\pi/2,\pi/2] $ so that $\cos{\theta_0} > 0$. Therefore, $N^2<0$ is equivalent to 

\begin{equation}\label{eq:oldgrav}
    x<-|\cot{\theta_0}| \mathrm{sgn}(b_r).
\end{equation}

\begin{figure}[b!]
\centering
\includegraphics[width=0.9\linewidth]{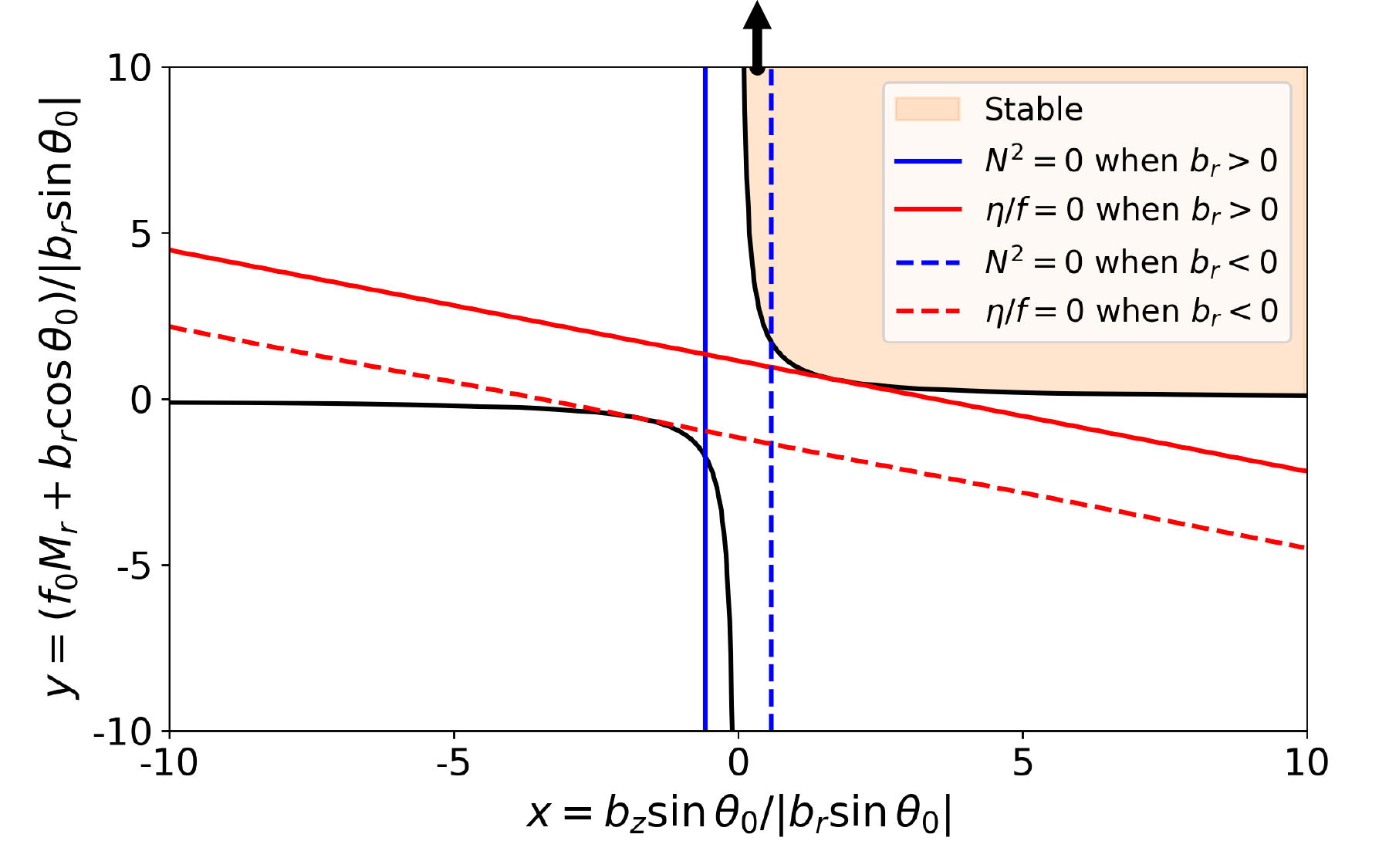}
\caption{Instability diagram for $\theta_0 = 60^\circ$ (other latitudes have similar results). Blue lines indicate $N^2=0$, and red lines indicate $\eta/f=0$, with solid lines indicating the results for $b_r>0$ and dashed lines indicating the results for $b_r<0$. The black lines indicate $xy=1$. The shading indicates the stable region. The black dot indicates the parameters for simulation $s_{S}$. Note that for $s_{S}$, the $y$-axis value is $1.15 \times 10^4$, outside the $y$-values shown in the plot.\label{figS1:N2}}
\end{figure}

The blue lines mark Equation~\ref{eq:oldgrav} in Figure~\ref{figS1:N2} when $\theta_0=60^\circ$. As indicated in the plot, when $b_r<0$, $N^2<0$ is not a sufficient condition for instability as the system can remain stable even when $N^2<0$, which is consistent with results shown in the numerical simulation $s_{S}$.

We next consider the case $\eta/f<0$, where, for easier comparison with previous work \citep[e.g.,][]{itano_symmetric_2009}, we neglect the geometric terms so that $f_0=2\Omega$ and $M_r = f_0 + \partial \overline{u}/\partial r$. Note that 

\begin{equation}
    \frac{1}{R}\frac{\partial \overline{u}}{\partial \theta} = \pz{\overline{u}} \cos{\theta_0} - \pr{\overline{u}} \sin{\theta_0} = M_z \cos{\theta_0} - (M_r - f_0)\sin{\theta_0}.
\end{equation}

As a result,

\begin{eqnarray}
\frac{\eta}{f} &=& \frac{f_0 \sin{\theta_0}-\frac{1}{R}\frac{\partial \overline{u}}{\partial \theta_0}}{f_0 \sin{\theta_0}} \nonumber \\
&=& \frac{M_r \sin{\theta_0} - M_z \cos{\theta_0}}{f_0 \sin{\theta_0}} \nonumber \\
&=& \frac{M_r \sin{\theta_0} - \frac{b_r \sin{\theta_0} - b_z\cos{\theta_0}}{f_0} \cos{\theta_0}}{f_0 \sin{\theta_0}} \nonumber \\
&=& \frac{f_0 M_r - b_r \cos{\theta_0} + b_z \sin{\theta_0} \cot^2{\theta_0} }{f_0^2} \nonumber \\
&=& \frac{|b_r \sin{\theta_0}|}{f_0^2} \left[ \frac{f_0 M_r + b_r \cos{\theta_0}}{|b_r \sin{\theta_0}|} + \frac{b_z \sin{\theta_0} \cot^2{\theta_0}}{|b_r \sin{\theta_0}|} - \frac{2b_r \cos{\theta_0}}{|b_r \sin{\theta_0}|}\right] \nonumber \\
&=& \frac{|b_r \sin{\theta_0}|}{f_0^2} \left[ y + \cot^2{\theta_0} x - 2 |\cot{\theta_0}| \mathrm{sgn}(b_r) \right].
\end{eqnarray}

\noindent where $y = (f_0 M_r + b_r \cos{\theta_0})/|b_r \sin{\theta_0}|$ is defined in Equation~\ref{eq:definexy}. Therefore, $\eta/f<0$ is equivalent to 

\begin{equation}\label{eq:oldiner}
    y < - \cot^2{\theta_0} x + 2 |\cot{\theta_0}| \mathrm{sgn}(b_r).
\end{equation}

The separating line described by Equation~\ref{eq:oldiner} is always tangential to the hyperbolic curve, $xy=1$ (see the red lines in Figure~\ref{figS1:N2} for the case with $\theta_0=60^\circ$). $\eta/f<0$ is therefore always a sufficient (although not necessary) condition for instability.

\section{Details of the simulation setup}\label{app:simulation}

The perturbation equations~\ref{eqa1:u}--\ref{eqa5:cont} form the system that we numerically integrate in time with Dedalus \citep{burns2020dedalus}, where we choose the domain scale of the numerical simulations as the length scale $L$. We neglect the last term in Equation~\ref{eqa5:cont}, consistent with the scaling analysis in \ref{app:dispersion}. To search for plane-wave solutions in an infinitely large domain, we apply a Cartesian coordinate system with double-periodic boundary conditions. The equations are solved on a grid with $256 \times 256$ grid points.

The nonlinear effects may induce turbulence in the system before the most unstable mode becomes dominant when simulations are initialized with white noise. To address this issue and emphasize the initial exponential growth phase, we integrate the linearized equations (Equations~\ref{eqa1:u}--\ref{eqa5:cont} without the advection terms, $\hat{v}\partial/\partial \hat{r}$ and $\hat{w}\partial/\partial \hat{z}$) with white noise in the $\hat{b}$ field as the initial condition for at least $10/\hat{\sigma}_m$, where $\hat{\sigma}_m$ is the expected maximum growth rate, until the most unstable mode dominates. Subsequently, we use these modes, albeit with reduced amplitude, as the initial conditions for nonlinear simulations in each scenario (first column in Figure~\ref{fig5:snapshots}). Only the results of these nonlinear simulations are presented in this paper. In $s_{S}$, we adopt the same initial conditions as in $s_{G}$ for the nonlinear simulation, representing both stable and unstable scenarios with $\mathrm{Ro} \ll 1$.

\begin{figure}[h!]
\centering
\includegraphics[width=1.0\linewidth]{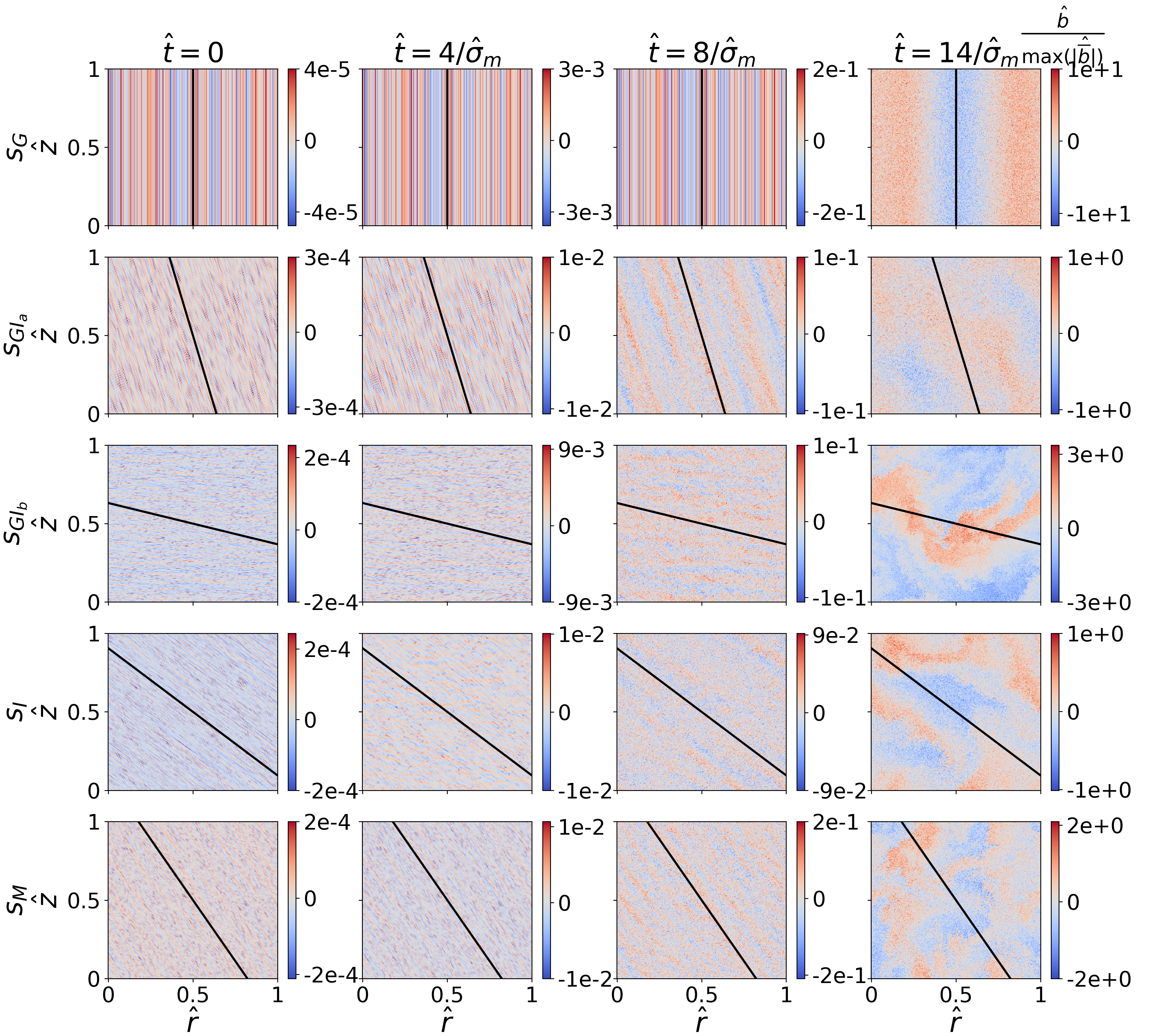}
\caption{Same as Figure~\ref{fig5:snapshots}, but for simulations with no diffusivity and viscosity. \label{figS2}}
\end{figure}

To address the problem of grid-scale noise in the non-linear simulations without viscosity and diffusivity (Figure~\ref{figS2}), we employ sub-grid parameterizations to represent the effects of sub-grid-scale eddy mixing via a flow-dependent eddy diffusivity ($\kappa_e$) and viscosity ($\nu_e$), computed following the parameterization of \cite{leith1996stochastic}. The Leith sub-grid parameterization is based on the theory of 2-D turbulence, where enstrophy cascades toward smaller scales and is ultimately dissipated at the grid scale. As a result, the eddy diffusivity and viscosity are parameterized as follows:

\begin{equation}
    \nu_e = \kappa_e = \left( \frac{K_{Leith}}{\pi} \right)^3 L_{grid}^3 \sqrt{\left[ \frac{\partial}{\partial \hat{r}} \left( \frac{\partial \hat{v}}{\partial \hat{z}} - \frac{\partial \hat{w}}{\partial \hat{r}} \right) \right]^2 + \left[ \frac{\partial}{\partial \hat{z}}  \left( \frac{\partial \hat{v}}{\partial \hat{z}} - \frac{\partial \hat{w}}{\partial \hat{r}} \right) \right]^2},
\end{equation}

\noindent where $L_{grid}$ is the grid scale and $K_{Leith}$ is a proportionality parameter, chosen as $K_{Leith}=1$ in our simulations. In addition to simulations with Leith viscosity and diffusivity, we also perform simulations without any explicit  viscosity and diffusivity, and compare their results (see Figures~\ref{fig4:timeseries_nonlinear}, \ref{fig5:snapshots}, and \ref{figS2}). During the initial, approximately exponential, growth stage, the two simulation sets behave similarly albeit with a somewhat reduced growth rate in the simulations with eddy viscosity and diffusivity. Once the simulations become strongly nonlinear, the simulations without Leith viscosity and diffusivity develop excessive noise at small scales, although the large-scale results remain qualitatively similar over the time-scale considered in our simulations (Figure~\ref{figS2}).

\newpage

\newpage

\section*{Open Research Section}
The Dedalus \citep{burns2020dedalus} code, available at \href{https://dedalus-project.org}{dedalus-project.org}, was used to perform numerical simulations in this study. The data that support the findings of this study are openly available on Zenodo at \citet{zeng2024simulation}.

\acknowledgments
We thank Wanying Kang, Noboru Nakamura, Hao Fu, Yixiao Zhang, and an anonymous reviewer for helpful discussions and comments.

\bibliography{SI.bib}

\end{document}